 \documentclass[12pt,epsf]{article}  
\usepackage{graphicx,amsmath}

\usepackage{amsmath,amsfonts,latexsym,amssymb}
\usepackage{verbatim,amsthm,curves,graphics}
\usepackage{mathrsfs}
\usepackage[]{inputenc}
\usepackage[T1]{fontenc}
\usepackage{hyperref}
\usepackage{color}
\usepackage{lmodern}
\usepackage[toc,page]{appendix}
\usepackage{fullpage}
\usepackage{authblk}

\def\ben{\begin{equation}}
\def\een{\end{equation}}
\def\bena{\begin{eqnarray}}
\def\eena{\end{eqnarray}}
\def\half{{1\over 2}}
\def\quater{{1 \over 4}}

\def\f(#1/#2){\frac{#1}{#2}} 
\def\Frac(#1/#2){\left(\frac{#1}{#2}\right)} 
\def\chris(#1-#2-#3){{\mit \Gamma}^{#1}{}_{{#2}{#3}} }
\def\tilchris(#1-#2-#3){\tilde{{\mit \Gamma}}^{#1}{}_{{#2}{#3}}}
\def\hatchris(#1-#2-#3){\hat{{\mit \Gamma}}^{#1}{}_{{#2}{#3}}}

\newcommand{\non}{\nonumber}
\theoremstyle{definition}

\newcommand{\dd}{{\rm d}}
 \newcommand{\cH}{\mathcal H}
\newcommand{\cI}{{\mathcal I}}
\newcommand{\cD}{{\mathcal D}}
\newcommand{\cM}{{\mathcal M}}
\newcommand{\cB}{{\mathcal B}}
\newcommand{\cC}{{\mathcal C}}

\newcommand{\mr}{{\mathbb R}}

\newcommand{\bS}{{\mathbb S}}
\newcommand{\bV}{{\mathbb V}}
\newcommand{\bT}{{\mathbb T}}

\renewcommand{\pounds}{\mathscr{L}}

\begin{document}

\author{Stefan Hollands}
\affil{\small Institute for Theoretical Physics, Leipzig University, Br\" uderstrasse 16, 04103 Leipzig, and MPI-MiS, Inselstrasse 22, 04103, Leipzig, Germany, stefan.hollands@uni-leipzig.de}
\author{Akihiro Ishibashi}
\affil{\small Department of Physics and Research Institute for Science and Technology, Kindai University, Higashi-Osaka, Osaka 577-8502, Japan, akihiro@phys.kindai.ac.jp}
\author{Harvey S. Reall}
\affil{\small Department of Applied Mathematics and Theoretical Physics, University of Cambridge, Wilberforce Road, Cambridge CB3 0WA, United Kingdom, hsr1000@cam.ac.uk}

\title{A stationary black hole must be axisymmetric in effective field theory}

\maketitle

\begin{abstract}
The black hole rigidity theorem asserts that a rotating stationary black hole must be axisymmetric. This theorem holds for General Relativity with suitable matter fields, in four or more dimensions. We show that the theorem can be extended to any diffeomorphism invariant theory of vacuum gravity, assuming that this is interpreted in the sense of effective field theory, with coupling constants determined in terms of a ``UV scale'', and that the black hole solution can locally be expanded as a power series in this scale. 
\end{abstract}

\section{Introduction}

Consider a stationary black hole spacetime of dimension $d \ge 4$.  The orbits of the asymptotically timelike Killing vector field (KVF) $t^a$ must leave the horizon invariant and are either 
everywhere tangent to the null-generators of the horizon, or not. It is known in standard Einstein gravity coupled to a wide range of standard matter models that in the first case, 
the metric and matter fields are actually static \cite{sudarsky,Israel67,Israel68,bunting,gibbons,gibbons1,Gibbons02a,ruback,Rogatko:2003kj}. In the other case, the black hole horizon is said to be rotating. In such a case, it is known, again for a fairly general class of standard matter models but assuming that the spacetime is real analytic and non-degenerate, that there necessarily exists another KVF $\chi^a$ that is tangent to the null-generators of the horizon; furthermore it is possible to show that $t^a = \chi^a + \sum_j \Omega_j^{} \psi_j^a$, where the $\psi_j^a$ are commuting KVFs each of which has closed orbits with period $2\pi$. Thus, the black hole is necessarily stationary {\it and} axi-symmetric. This theorem is originally due to Hawking \cite{H72,he}, who considered $d=4$ dimensions. Later improvements include \cite{frw,Racz00,RW96,RW92} (partially eliminating the analyticity assumption) as well as 
\cite{klainerman} (eliminating the analyticity- but under a smallness assumption). For higher dimensions $d \ge 4$ see \cite{moncrief,hiw,hi}\footnote{See also \cite{mi1,mi2,mi3} for closely related work on 
Cauchy horizons with closed generators.}.
These results are often called ``rigidity theorem'', because they imply among other things that the black hole must be rotating rigidly with respect to infinity with angular velocities $\Omega_j$.

In $d=4$, the rigidity theorem is an important stepping stone for the proof of the uniqueness -- or ``no hair'' -- theorems \cite{Bunting83,Carter71,Robinson75,Mazur82,costa} for Kerr-Newman black holes. 
Even though no uniqueness theorems of comparable strength are known in $d>4$, or in a number of Einstein-matter theories even in $d=4$, the rigidity theorem is of major structural importance not least because it is a prerequisite for the zeroth- and first laws of black hole mechanics. It is therefore natural to ask whether the rigidity theorem remains valid for example in the presence of higher-derivative terms in the action as expected from an effective field theory (EFT) perspective. 

In this paper, we will prove an extension of the rigidity theorem in dimensions $d \ge 4$ for general, local, covariant purely gravitational EFTs extending standard Einstein general relativity. 
Our approach is to order the terms in the EFT action/equations of motion by the numbers of derivatives that they contain and to study, so to speak, the effects of these terms
to increasing accuracy. Each term in the action is multiplied by a suitable power of some length scale 
$\ell$ that one may think of as a cutoff scale for UV physics if desired. For example, the standard Einstein-Hilbert Lagrangian $R$
has two derivatives; the next possible terms would be linear combinations of quadratic curvature invariants with four derivatives, 
$R^2, R_{ab}R^{ab}$ and $R_{abcd}R^{abcd}$, each multiplied by $\ell^2$, and so on. Roughly speaking, 
the EFT approach is to restrict attention to solutions varying over a typical length scale $L$ such that $\ell/L \ll 1$, and this means, 
roughly speaking, that for such solutions the higher curvature terms are always smaller than the leading Einstein Hilbert term in the action. More precisely, one may say (see def. 4.1 of \cite{hkr}) 
that the EFT condition is valid near the horizon, $\cH$, if considering a 1-parameter family of solutions to the theory with parameter $L$ (in our case thought of roughly as the size of the black hole), 
any quantity of dimension $n$ built from coordinate components of the metric in a suitable class of coordinate systems will remain bounded by $C_n/L^n$ in absolute value. Then, if $\ell \ll L$, the 
higher derivative term of dimension $D$ will, intuitively, locally make a very small correction $(\ell/L)^D$ to the solution. Thinking of the length scale $L$ as fixed instead and making $\ell$ small, 
this motivates that we should ask whether the expansion coefficients in $\ell$ of a family of solutions $g_{ab}(\ell,x)$ labelled by the UV length scale $\ell$ 
locally satisfy the rigidity theorem order by order. This is what we shall actually do in this paper. 

The result of this analysis in the rotating case is thm. 3 (and its local version thm. 1) and it assumes, just as the rigidity theorem in ordinary Einstein gravity, 
that the family of metrics is real analytic in the domain of outer communication and non-degenerate. Furthermore, it appears that we additionally need to assume a certain 
genericity requirement in $d>4$ which however does not impose a major restriction physically.  Even though thm. 3 is an order-by-order statement for the expansion in $\ell$ -- and thus 
already a good approximation in view of the EFT hypothesis near any horizon cross section -- 
if the metric was known to be jointly analytic in $\ell$ and $x$, it is plausible that the rigidity theorem, 
i.e. existence of the further KVFs $\chi^a, \psi_j^a$ would actually hold for finite $\ell$ sufficiently close to zero. 

The proof of  thms. 1 and 3 is inductive, with the induction ascending in the power of $\ell$. 
For vanishing $\ell$, the usual rigidity theorems in standard Einstein gravity of course apply. To make the induction step, we follow the ideas 
used in standard Einstein gravity \cite{hiw} up to a point. However, we cannot at higher orders in $\ell$ parallel a key step employed in \cite{hiw} which is using the Raychaudhuri equation, because it 
gives insufficient information in the presence of higher derivative terms in the action. As is well known, the Raychaudhuri equation plays a key role in the proof of the area- and singularity -- and also the rigidity theorem -- in ordinary Einstein gravity (see e.g. \cite{he,wald}) and its main use comes about because the stress energy tensor ordinarily comes in with a definite sign. 
By contrast, in our higher derivative theory, the higher order derivative terms in the equation of motion produce terms in the 
Raychaudhuri equation which are not sign-definite in general.  However, we are able, within our inductive 
scheme, to replace the arguments normally based on the Raychaudhuri equation and the horizon area with an argument involving an entropy current-density in EFTs recently analyzed in \cite{hkr} (for previous works on such entropy current-densities see \cite{Iyer94, Jacobson1995, Wall2015, Bhat17,Bhat19,Bhat21}). Furthermore, for higher derivative theories, the treatment of the 
rotational Killing fields seems more subtle than for Einstein gravity.

A corollary of our proof is that the surface gravity, $\kappa$, which can be defined thanks to the existence of the additional KVF $\chi^a$ tangent and normal to the horizon, 
is constant, i.e. that the zeroth law of black hole mechanics holds. It is interesting to note that if one {\it assumes} the existence of $\chi^a$, the zeroth law can be demonstrated in the EFTs that we are considering by an independent argument \cite{Bhat22,gosh}.

\medskip

This paper is organized as follows. In sec. \ref{sec:1} we present in detail our assumptions and recall Gaussian null coordinates and related constructions required in 
the proof of the local rigidity theorem thm. 1, which is presented in sec. \ref{sec:rot}. In sec. \ref{sec:nonrot} we analyze for completeness the situation regarding KVFs for non-rotating horizons summarized in thm. 2 , and in sec. \ref{sec:analytic}, we present our main result thm. 3. Some technical material is relegated to various appendices. Our conventions and notations are the same as in \cite{wald}. Lower case Roman indices $a,b,c, \dots$ are abstract spacetime indices whereas Greek indices $\mu,\nu,\sigma, \dots$ refer to specific spacetime coordinates depending on the context. Upper case Roman indices $A,B,C, \dots$ refer to coordinates on the horizon cross section, $\cC$. We work in units such that $16\pi G=1$.
 
 \pagebreak

\section{Setup} 
\label{sec:1}

\subsection{Standing assumptions}
We consider a generic covariant parity even\footnote{This assumption is made only for simplicity and there is no difficulty in principle to generalize our proofs to theories with parity odd terms.} gravitational theory describing corrections to Einstein gravity. Such a theory is 
described by an action of the form
\ben
\label{theory}
I[g] = \int_{\cM} (R + \ell^2 L_4 + \ell^4 L_6 + \dots + \ell^{2D-2} L_{2D})\sqrt{-g} \dd^d x
\een
where the $L_{2j}$ are covariant local functionals of the metric containing $2j$ derivatives\footnote{If we drop the parity even requirement, we can use the 
volume element $\epsilon_{a_1 \dots a_d}$ to obtain terms with an odd number of derivatives in odd $d$.}. By the Thomas replacement theorem \cite{Iyer94}, each $L_{2j}$ 
is therefore a contraction of the (inverse) metric with a tensor product of $\nabla_{a_1} \dots \nabla_{a_r} R_{abcd}$. We could add a cosmological constant term $L_0=-2\Lambda$, which would result 
in an obvious change in the asymptotic conditions on the metric, but not in a major change in our proofs. We will briefly comment on this in remark 1) below thm. 3.

The Euler Lagrange (Einstein-) equations for \eqref{theory} can  be written in the schematic form 
\ben
\label{Ein:eq}
G_{ab} =  \ell^2 H_{4 \, ab} + \dots + \ell^{2D-2} H_{2D \, ab}
\een
with local covariant tensors $H_{2j\, ab}$ containing $2j$ derivatives of the metric. 
The standing assumptions on the solutions to \eqref{Ein:eq} considered in this paper are:
\begin{enumerate}
\item 
We have a 1-parameter family of $d$-dimensional ($d \ge 4$), stationary, asymptotically Minkowskian  solutions $(\cM, g_{ab})$ to \eqref{Ein:eq} with asymptotically timelike KVF $t^a$
with complete orbits, see defs. 2.1, 2.2 of \cite{CW94} for the precise asymptotic and causality conditions. Both $t^a, g_{ab}$ are functions of the parameter $\ell$ and $(\cM, g_{ab})$ contains a black hole. We require that the manifold structure of $\cM$ is independent of $\ell$ and, as a gauge condition, we require that the location of the (future) horizon\footnote{
\label{foot:1}
It is defined as the future boundary of the 
domain of outer communication. More precisely, as in \cite{CW94}, we assume that $\cM$ contains an acausal hypersurface $\Sigma$ with possibly several 
asymptotic ends $\Sigma_1, \Sigma_2, \dots$ each
$\cong {\mathbb R}^{d-1} \setminus B_R$. The slice must satisfy either def. 2.2(a,b) of \cite{CW94}, which precludes it from ``not reaching the event horizon''.
Then the domain of outer communication $\cD$ with respect to the end $\Sigma_1$ is 
defined as the causal completion of $\cup_{t \in {\mathbb R}} \phi_t[\Sigma_1]$ where $\phi_\tau$ is the flow of $t^a$ and the future/past horizon is defined as 
$\cH^\pm = \partial \cD \cap I^\pm(\cD)$.}, $\cH$, is independent of $\ell$. We also require that $\cH$ is smooth.
\item 
$g_{ab}(\ell, x)$ and $t^a(\ell,x)$ are jointly smooth in $(\ell, x)$, meaning that we have an asymptotic expansion of the form 
\ben
\label{asymptex}
g_{ab}(\ell, x) = \sum_{n = 0}^{2M} \ell^n g_{ab}^{(n)}(x) + O(\ell^{2M+2})(x)
\een 
where each $g_{ab}^{(n)}$ is smooth on $\cM$ and where $M$ can be as large as we like, and similarly for $t^a$. 
We assume w.l.o.g. that only even powers of $\ell$ appear, the same goes for all similar expansions below. Here and in the following, 
$O(\ell^n)$ denotes a term such that for each coordinate neighborhood $\mathcal{U}$ of $\cH$ with compact closure and each $k \ge 0$, there exists a sufficiently small $\ell_0=\ell_0(\mathcal{U},k)$ and a constant $c_{n} = c_n(\mathcal{U},k)$ such that $|\partial_{\mu_1} \dots \partial_{\mu_k} O(\ell^n)(x)| \le c_{n} \ell^n$ for all $|\ell| \le \ell_0$, all $x \in \mathcal{U}$.

\item
We assume that $\cH$ 
has topology $\cH = {\mr} \times \cC$, where $\cC$ is compact and that $\cH$ is non-degenerate for $\ell = 0$ [for the precise definition see below eq. \eqref{kappadef}]. 
\end{enumerate}

We note that the asymptotic expansion in $\ell$ postulated in item 2) is not required to be uniform in $x$, e.g. we allow that the metric for finite $\ell$ could deviate from the $\ell=0$ solution in standard Einstein gravity by an ever increasing amount as time goes to infinity, no matter how small $\ell$. In other words, we allow the corrections from the higher derivative terms, while locally small, to pile up in an unbounded manner over asymptotically large times. In a sense, we are therefore allowing secular effects. 

While in the case of $d=4$ dimensions, the above requirements will be sufficient for the proof of thm. 1, our method of analysis appears to necessitate a further ``genericity'' assumption in higher dimensions $d>4$, unless the horizon is non-rotating. Since the analysis of both cases is rather different anyhow, we shall distinguish them in the following:

\begin{enumerate}
    \item[I)] Rotating Case: $t^a$ is {\em not} tangent to the null generators of $\cH$ for sufficiently small $|\ell|$.
    \item[II)] Nonrotating Case: $t^a$ is tangent to the null generators up to arbitrary order in $\ell$. 
\end{enumerate}

There is of course also the possibility that $t^a$ is tangent to the null generators of $\cH$ only up to a finite order in $\ell$. The treatment of this case would require a combination of the 
methods in cases I) and II), depending on the order in $\ell$ in the induction procedure. Since it is only case I) that should be considered generic anyhow, and since the analysis would be rather repetitive, 
we will not give it here. In case II), it could also in principle happen that $t^a$ is not tangent to the null generators for a sequence $\{ \ell_n \}$ tending to zero if $t^a, g_{ab}$ and/or the manifold 
structure of $\cM$ is not analytic, in which case the terminology `nonrotating' is misleading. However, this case will not be relevant for our analysis since we will only obtain results order by order in $\ell$ for thm. 1 or assume analyticity for thm. 3. The rotating case is treated in sec. \ref{sec:rot} whereas the non-rotating case is 
treated in sec. \ref{sec:nonrot}. 

In order to state the ``genericity'' assumption in the rotating case, 
we need to recall -- and will demonstrate again below -- that for $\ell = 0$, i.e. Einstein gravity, already the above assumptions 1)-3) 
imply that the projection of the flow generated by $t^a|_{\ell=0}$ to any cross section $\cC$ of $\cH$ is a Killing vector field, $S
^a$, of the metric $g_{ab}|_{\ell=0}$
restricted to $\cC$. Let $\{\hat \phi_\tau \ : \ \tau \in \mathbb{R} \}$ be the flow of $S^a$ on $\cC$. It is an abelian subgroup, $\mathcal A$, of the isometry group of the compact Riemannian manifold $\cC$ (with the metric induced from $g_{ab}|_{\ell=0}$). Its closure, $\mathcal G=\overline{\mathcal A}$, therefore is an abelian compact Lie-group, hence isomorphic to a torus ${\mathcal G} = \mathbb{T}^N$ for some $N \ge 1$. The $N$ generators of this torus correspond to $N$
KVFs of $\cC$, $\psi^a_{1}, \dots, \psi^a_{N}$, each generating a flow of isometries with period $2\pi$, and we have, on $\cC$,
\ben
\label{S:lin}
S^a = \sum_{j=1}^N \Omega_{j}^{} \psi^a_{j}.
\een

\begin{enumerate}
\item[4.] (Genericity)
We assume that the flow of $S^a$ generates the {\it full} isometry 
group of the cross sections $\cC$ of $\cH$ for the restriction of the metric $g_{ab}|_{\ell=0}$.
In particular, the isometry group must be the abelian group ${\mathbb T}^N$.
\end{enumerate}

Note that the genericity property is trivially fulfilled in $d=4$ dimensions since in that case, the stationary black holes in question are provided by the Kerr-family, 
which has only one rotational KVF. The Myers-Perry black holes \cite{mp} in $d$ dimensions have isometry group ${\mathbb R} \times {\mathbb T}^N$, 
where $N=\lfloor (d-1)/2 \rfloor$, unless some of the spin parameters $a_j$ happen to vanish. The genericity requirement imposes that the orbit of $S^a$
on a horizon cross section is dense in ${\mathbb T}^N$, and this will be the case if the $a_j$ are such that all non-trivial ratios $\Omega_i/\Omega_j$ are irrational numbers. 
Thus, 4) amounts to a genericity requirement on the values of the spin parameters $a_j$ which is satisfied for almost all values of these parameters because 
the irrational numbers are dense in the real numbers.

\subsection{Gaussian null coordinates (GNCs)}

We begin by picking an arbitrary compact cross section $\cC$ of $\cH$ and flow it with the 1-parameter group $\phi_\tau$ of isometries generated by the KVF $t^a$ by an amount $v$ and set $\cC(v):=\phi_v[\cC]$. Using the global structure of spacetime expressed in assumption 1) and arguments similar to those given in the proof of prop. 4.1 of \cite{CW94}, 
we may assume without loss of generality that $\cC$ has been chosen so that each orbit of $t^a$
on $\cH$ intersects $\cC$ precisely once, so $t^a$ is everywhere transverse to each $\cC(v)$.
On each $\cC(v)$ we can therefore decompose
\ben
t^a = k^a + s^a
\een
where $s^a$ is tangent to each $\cC(v)$, not identically zero on $\cC(v)$ for rotating horizons, and $k^a$ is tangent and normal to $\cH$ and nowhere vanishing, see
fig. 1.
\begin{figure}[h!]
\begin{center}
  \includegraphics[width=0.7\textwidth,]{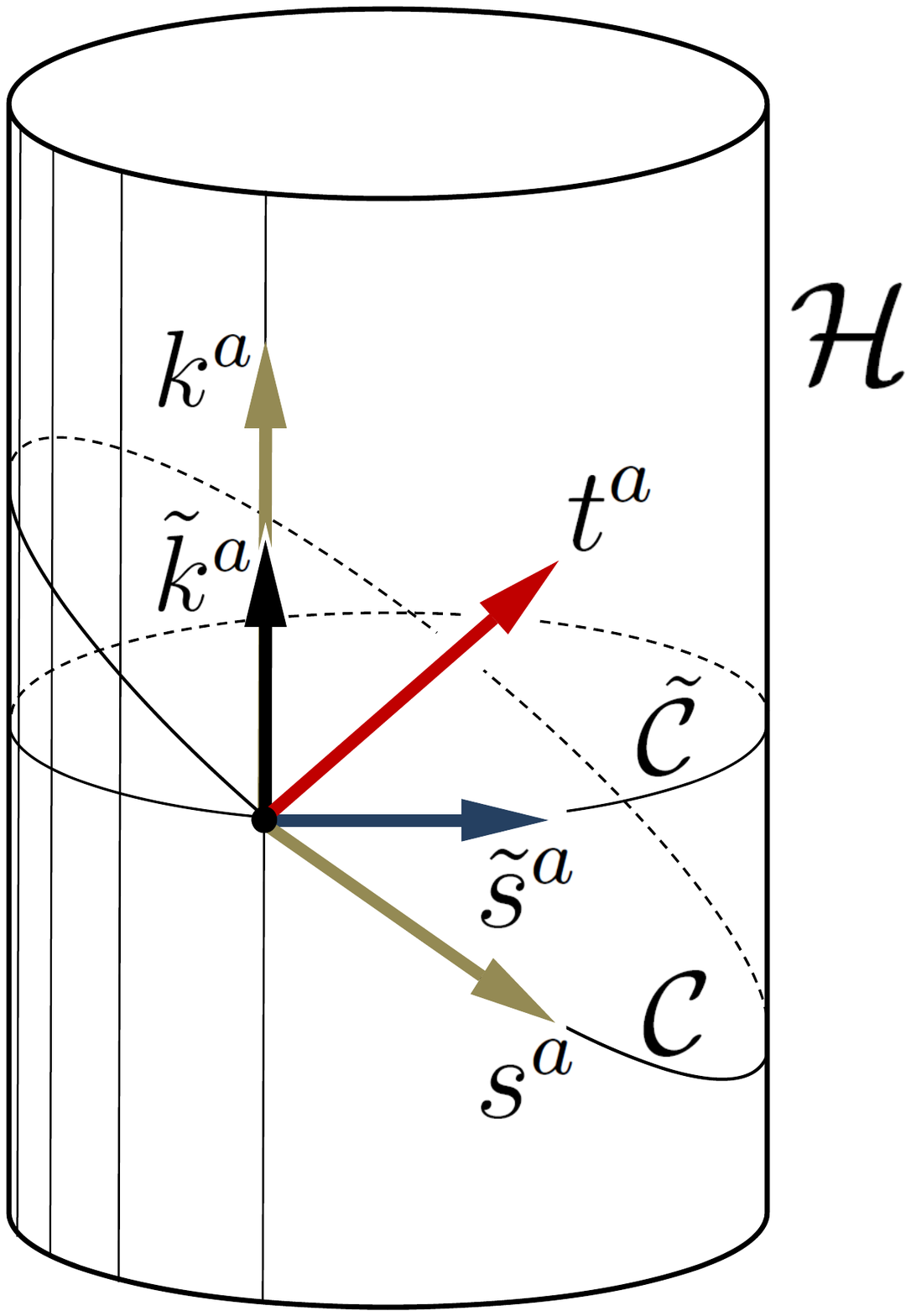}
  \end{center}
  \label{fig:1}
  \caption{Illustration of $t^a = k^a + s^a$ for a given cut $\cC$. If a different cut $\tilde \cC$ is chosen, we 
  get a correspondingly different $\tilde k^a$ and $\tilde s^a$.}
\end{figure}
By construction, we have 
\ben
\label{com}
\pounds_t k^a = \pounds_t s^a = \pounds_k s^a = 0, \quad \pounds_t v = 1
\een
on $\cH$. The family of cross section $\cC(v)$ defines a foliation of $\cH$ which we now use to set up an adapted Gaussian null coordinate (GNC) system in an open neighborhood of $\cH$; see app. A for further explanations about GNCs. To this end, we consider at each 
point of $\cH$ a second null vector $l^a$ normalized such that $l^a k_a = 1$ and such that $l^a$ is perpendicular to the corresponding cut $\cC(v)$. We extend $l^a$
off of $\cH$ imposing the geodesic equation $l^a \nabla_a l^b = 0$ and let $r$ be an affine parameter on each such geodesic such that $r=0$ on $\cH$. 
Finally, we may locally pick a coordinate system $(x^A)$ on  $\cC$, which is transported off of $\cC$ demanding that $\pounds_k x^A = \pounds_l x^A = 0$ where defined. 
Then, in the coordinates $(v,r,x^A)$, the metric takes the Gaussian Null Form:
\ben\label{gnc1}
g=2\dd v(\dd r - r\alpha \dd v  -r \beta_A \dd x^A) + \gamma_{AB} \dd x^A \dd x^B
\een
By construction we have
\ben
\label{GNCvectors}
k^a = \left( \frac{\partial}{\partial v} \right)^a , \quad 
l^a = \left( \frac{\partial}{\partial r} \right)^a, \quad
s^a = s^A \left( \frac{\partial}{\partial x^A} \right)^a,
\een
and we set 
\ben
\gamma_{ab} = \gamma_{AB} (\dd x^A)_a (\dd x^B)_b, \quad 
\beta_{a} = \beta_{A} (\dd x^A)_a 
\een
Even though we will of course need more than one coordinate chart $(x^A)$ to cover $\cC$,  
these coordinate charts can be patched together so that the above tensor fields  are defined globally and invariantly 
in an open neighborhood of $\cH$, depending only on the initial choice of $\cC$. 
By construction, we have 
\ben
\pounds_t \gamma_{ab} = \pounds_t \beta_a = \pounds_t \alpha = \pounds_t r = 0.
\een
In particular,  since $[k,s]^a = 0$ on $\cH$ by \eqref{com}, it follows that the GNC components of $s^a$ are independent of $v$ on $\cH$. Also, by construction,
we have
\ben
k^a \nabla_a k^b = \alpha k^b \quad \text{on $\cH$.}
\een
We can think of the tensors $\alpha, \beta_a, \gamma_{ab}$ as living on the foliation $\cC(v,r)$ of surfaces of constant $r,v$, and it will be useful to define a corresponding intrinsic covariant 
derivative operator and projections. For this purpose, we set $p^a{}_b = (\partial_A)^a (\dd x^A)_b$, and $q^{ab} = (\gamma^{-1})^{AB} (\partial_A)^a (\partial_B)^b$. Then $q^a{}_b$ is the orthogonal 
(with respect to $g_{ab}$) projector onto $T\cC(v,r)$, and we have $\gamma_{ab}q^{bc} = p^c{}_a$. $p^a_b$ is another projection onto $T\cC(v,r)$
characterized by $p^a{}_b l^b = 0 = p^a{}_bk^b$. Note that $p^a{}_b$ is not an {\it orthogonal} projection where $r\beta_a$ is non-vanishing and therefore not equal to $q^a{}_b$ nor to $\gamma^a{}_b$ at such points. But on $\cH$, i.e. for $r=0$, the quantities $q^a{}_b, \gamma^a{}_b, p^a{}_b$ all coincide. 
By construction, we have 
\ben
p^a{}_b \beta_a = \beta_b, \quad  p^a{}_b p^c{}_d \gamma_{ac} = \gamma_{bd}
\een
where these quantities are defined, and on $\cH$ we could replace $p^a{}_b$ in these expressions by $q^a{}_b$.

For a covariant tensor field $T_{a_1 \dots a_r}$ we define 
\ben
\label{Ddef}
D_b T_{a_1 \dots a_r} := q_b{}^c q_{a_1}{}^{d_1} \cdots q_{a_r}{}^{d_r} \nabla_c T_{d_1 \dots d_r}, 
\een
and we will denote by $R[\gamma]_{abc}{}^d$ the curvature of $D_c$ which is an intrinsically defined tensor field on each $\cC(v,r)$ (i.e. coincides with its projection via $q^a{}_b$). The contractions of $R_{ab}$ into $l^a, k^a, p^b{}_a$ as expressed in terms of $D_a, \gamma_{ab}, \beta_a, \alpha$ etc. are given in app. A. 

\section{Rotating case}
\label{sec:rot}

Assume that we are in the rotating case I).
We will give in this section a proof that there exists a KVF $\chi^a$ tangent to the null generators of $\cH$ in the sense that $\chi^a$ Lie derives $g_{ab}$ modulo terms of order $O(\ell^{n})$ where $n$ can be chosen as large as we like, and modulo terms that vanish to arbitrarily high order in any coordinate transverse to $\cH$. $\chi^a$ is constructed such that it commutes with $t^a$ and on 
$\cH$ satisfies $\chi^a \nabla_a \chi^b = \kappa \chi^b$, where $\kappa > 0$ is constant on $\cH$. 
If the solution is jointly real analytic in 
$(x,\ell)$, we will argue in the next section that $\chi^a$ has an analytic continuation to the entire domain of outer communication which Lie-derives $g_{ab}$ exactly, i.e. without any error terms. 

 The basic idea is to define $\chi^a := k^a$ where $k^a$ is the vector field (VF) tangent to the horizon generators defined by fixing a cross section $\cC$ of $\cH$
 in our construction of GNCs, see \eqref{GNCvectors}. The cross section $\cC$ is arbitrary in our construction of GNCs and different choices will lead to different $k^a$. We are going to 
 find the right $\cC$ by demanding that for the $k^a$ defined by that $\cC$, we have 
$\alpha = \kappa =$ constant on $\cH$. A generic $\cC$ will not do for this purpose, 
so we will have to pass to a new $\tilde \cC$, to be determined. The determination of this $\tilde \cC$ will be 
made order by order in $\ell$. To organize the powers of $\ell$, we define the expansion coefficients in
\ben
\begin{split}
\alpha(x,\ell) \sim& \sum_{n=0}^{\infty} \ell^n \alpha^{(n)}(x)\\
\beta_a(x,\ell) \sim& \sum_{n=0}^{\infty} \ell^n \beta^{(n)}_a(x)\\
\gamma_{ab}(x,\ell) \sim& \sum_{n=0}^{\infty} \ell^n \gamma^{(n)}_{ab}(x) 
\end{split}
\een
where ``$\sim$'' means an asymptotic expansion in the sense described below \eqref{asymptex}.
Similar expansions are made for other tensor fields on $\cM$, and we note that due to the structure of the action \eqref{theory}, only even powers of $\ell$ can appear in such expansions.
The conditions $\pounds_t \gamma_{ab}(\ell,x)  = \pounds_t \beta_{a}(\ell,x)  = \pounds_t \alpha(\ell,x) = 0$ may then be expanded out in powers of $\ell$ to obtain conditions on the expansion coefficients.
We define $\kappa$ to be the average of $\alpha$ over $\cC$, 
\ben
\label{kappadef}
\kappa = \frac{1}{A[\cC]} \int_{\cC} \alpha \sqrt{\gamma} \dd^{d-2} x
\een
where $A[\cC]$ is the area defined w.r.t. $\gamma_{ab}$. Then we have $\kappa \sim \sum_{n=0}^{\infty} \ell^n \kappa^{(n)}$, and the precise form of assumption 3) is that 
$\kappa^{(0)}$ so defined is $>0$. We now make the following 

\medskip
\noindent
{\bf Inductive hypothesis at order $\ell^n$:} There exists a cross section $\cC$ with corresponding GNC system $(v,r,x^A)$ such that 
\ben
\label{ih}
\begin{split}
\pounds_l^m \Big( \pounds_k \gamma_{ab} \Big)_{r=0} &= O(\ell^{n+2})\\
\pounds_l^m \Big( \pounds_k \beta_{a} \Big)_{r=0} &= O(\ell^{n+2})\\
\pounds_l^m \Big( \pounds_k \alpha_{} \Big)_{r=0} &= O(\ell^{n+2})
\end{split}
\een
for any $m \ge 0$. Furthermore, $\alpha^{(j)} = \kappa^{(j)} = $ constant on $\cH$ for all $j \le n$ in that GNC system, and the expansion coefficients of $s^a$ satisfy
\ben
\label{ihsn}
s^{(j)a} \quad 
\text{are KVFs of $\gamma_{ab}|_{\ell=0}$ for all $j \le n$ on $\cC$.}
\een

\medskip
\noindent
{\bf Remarks:} 1) By the genericity assumption 4) of sec. 2.1, the $s^{(j)a}, j=0, \dots, n$ are {\it commuting} KVFs of the zeroth order in $\ell$ horizon metric $\gamma_{ab}|_{\ell=0}$, and we can write, on $\cC$,
\ben
s^a(\ell, x) = \sum_{i=1}^N \Omega_i^{}(\ell) \psi_i^a(x) + O(\ell^{n+2}).
\een
\noindent
2) The facts that $t^a$ Lie-derives $\alpha, \beta_a, \gamma_{ab}$ for all 
$\ell$ together with $t^a = k^a+s^a$ and \eqref{ih} give
\ben
\label{ihs}
\begin{split}
\pounds_l^m \Big( \pounds_s \gamma_{ab} \Big)_{r=0} &= O(\ell^{n+2})\\
\pounds_l^m \Big( \pounds_s \beta_{a} \Big)_{r=0} &= O(\ell^{n+2})\\
\pounds_l^m \Big( \pounds_s \alpha_{} \Big)_{r=0} &= O(\ell^{n+2})
\end{split}
\een

\subsection{Induction start $n=0$} 

The argument in this section has been presented in 
\cite{hiw} but we go through some of its steps as a preparation for the induction step to familiarize the reader with the basic logic of our argument. 
Let $\lambda$ be an affine parameter for the null geodesic generators of $\cH$ whose future directed tangent we denote by $n^a$. 
Then we have the corresponding expansion and shear $\theta, \sigma_{ab}$ on $\cH$, given by 
\ben
\theta = \gamma^{ab} \nabla_{(a} n_{b)}, \quad \sigma_{ab} = \gamma_a{}^c \gamma_b{}^d (\nabla_{(c} n_{d)} - \tfrac{1}{d-2} g_{cd} \nabla_e n^e). 
\een 
We consider the $v$-derivative of the area (with respect to $\gamma_{ab}$):
\ben
\label{dA}
\frac{\dd}{\dd v} A[\cC(v)]_{v=0} = \int_{\cC} \frac{\partial \lambda}{\partial v} \theta \sqrt{\gamma} \dd^{d-2}x
\een
For convenience, we will take $\lambda = 0$ on $\cC$. 
We know $\frac{\dd}{\dd v} A[\cC(v)]=0$, because the flow generated by $t^a$ is isometric by assumption 1) and because the area $A$ of a cut $\cC$ is covariant, i.e. only dependent on 
the metric structure and on $\cC$. The Raychaudhuri equation gives
\ben
\label{Ray}
\partial_\lambda \theta = -\sigma_{ab} \sigma^{ab} - \frac{1}{d-2} \theta^2 - R_{ab} n^a n^b
\een
and the Einstein equation \eqref{Ein:eq} gives $R_{ab} n^a n^b = O(\ell^2)$. By the same argument as for the area theorem
\cite{he} for the $\ell=0$ solution, we cannot have $\theta^{(0)}<0$ on $\cH$. Combining this statement with $\frac{\dd}{\dd v} A[\cC(v)]=0$
evaluated at order $O(\ell^0)$ therefore gives $\theta^{(0)} = 0$ on $\cH$. In view of \eqref{Ray} evaluated at $O(\ell^0)$, we conclude that $\sigma_{ab}^{(0)}=\theta^{(0)}_{}=0$ on $\cH$.

Thus, we have the first equation in \eqref{ih} for $n= m = 0$. Combining 
$\pounds_k \gamma^{(0)}_{ab} = 0$ on $\cH$ with $t^a = k^a + s^a$ and the fact that $\pounds_t \gamma_{ab} = 0$, this also shows that 
$S^a=s^a|_{\ell=0}$ Lie derives $\gamma^{(0)}_{ab}$, i.e. is a KVF.
These conclusions hold no matter how we chose the 
initial cut $\cC=\cC(0)$ to set up our GNC system.

However, the other equations in \eqref{ih} for $n=0$ are in general not satisfied for an arbitrarily chosen initial cut $\cC$. 
We now wish to find the appropriate new cut $\tilde \cC$ and thereby the appropriate decomposition $t^a = \tilde k^a + \tilde s^a$ (see fig. 1) and GNC system $(\tilde v, \tilde r, \tilde x^A)$ with 
associated tensors 
$\tilde \alpha, \tilde \beta_a, \tilde \gamma_{ab}$ satisfying \eqref{gnc1}.  For this, 
we consider the $vA$-component of the Ricci tensor on $\cH$, see \eqref{nAR}. Using that we now know $\pounds_k \gamma_{ab} = O(\ell^2)$, this gives 
\ben
\label{vA}
R_{bc} k^b  p^c{}_a = D_a \alpha - \frac{1}{2} \pounds_k \beta_a + O(\ell^2) \quad \text{on $\cH$.}
\een 
Using the Einstein equation on $\cH$, we have $R_{ac} k^a  p^c{}_b = O(\ell^2)$, so sending $\ell \to 0$ we find that $D_a \alpha^{(0)} - \frac{1}{2} \pounds_k \beta^{(0)}_a = 0$ on $\cH$,
which implies $D_a \alpha^{(0)} + \frac{1}{2} \pounds_{S} \beta^{(0)}_a = 0$ on $\cC$, where $S^a = s^a|_{\ell=0}$.
We see from this equation that if we can define a new cut $\tilde \cC$ with corresponding new $\tilde \alpha, \tilde \beta_a, \tilde \gamma_{ab}$ 
in such a way that $\tilde \alpha^{(0)}$ is constant on $\cH$, then $\pounds_{\tilde k} \tilde \beta^{(0)}_a = 0$, and we will have all the equations in \eqref{ih} for 
$n=0$ for $m=0$ for the tilde fields. Additionally, we will have learnt that $\tilde \alpha^{(0)} = \kappa{(0)} = $ constant.

Let us determine the conditions that the new cut $\tilde \cC$ would have to satisfy. It is clear that $\tilde k^a$ must be proportional to $k^a$, so it must be the case that
$\tilde k^a = f k^a$  for some positive function $f$ . Since $\pounds_t k^a = \pounds_t \tilde k^a = 0$, we must have $\pounds_t f = 0$ and therefore 
$\pounds_k f = -\pounds_s f$. Since
on $\cH$ we know that $k^a \nabla_a k^b = \alpha k^b$ and that $\tilde k^a \nabla_a \tilde k^b = \tilde \alpha \tilde k^b$. This means 
that $\alpha$ and $\tilde \alpha$ are related through $f$ by 
\ben
\label{11}
\pounds_k f + \alpha f = -\pounds_s f+ \alpha f = \tilde \alpha. 
\een
Demanding $\tilde \alpha = \kappa^{(0)}+O(\ell^2)$ 
means in view of the last relation 
that $f$ should be taken to be a solution of
\ben
\label{1}
 -\pounds_{S} f + \alpha^{(0)} f  = \kappa^{(0)} 
\een
using as before the notation $S^a = s^a|_{\ell=0}$.
Equality \eqref{1} provides a condition that must be necessarily be 
satisfied on $\cC$, cf. eq. (23) of \cite{hiw}. We must additionally have $\pounds_t \tilde v = 1$ on $\cH$ by \eqref{com}, and since $\tilde k^a = (\partial/\partial \tilde v)^a$,
\ben
\pounds_k \tilde v = \frac{1}{f} \pounds_{\tilde k} \tilde v =  \frac{1}{f}. 
\een
Using $t^a = k^a + s^a$ and taking $\ell=0$ 
for all the quantities shows that 
\ben
1-\pounds_{S} \tilde v = \frac{1}{f}. 
\een
Therefore $\tilde v$ must on $\cC$ satisfy the equation
\ben
\label{2}
\pounds_{S} \tilde v = 1 + \frac{1}{\kappa^{(0)}} \bigg(\pounds_{S} \log f - \alpha^{(0)} \bigg), 
\een
cf. eq. (38) of \cite{hiw}. It is easy to see that \eqref{1} and \eqref{2} have a solution on $\cC$ for $d=4$: In that case, by the horizon topology theorem \cite{he}, $\cC \cong {\mathbb S}^2$, so by standard results on isometric actions of spheres, the orbits of $S^a$ must close after a certain period $2\pi/\Omega$. It is then easy to see from this fact that 
 \eqref{1} and \eqref{2} have a solution on $\cC$. In $d>4$, even though $S^a$ need not have closed orbits on $\cC$, \eqref{1} and \eqref{2} have a solution on $\cC$ by lemmas 1 and 2 of \cite{hiw} since we are assuming to be in the non-degenerate case, $\kappa^{(0)}>0$. 

This gives $\tilde v$ as a function on $\cC$, and then we extend it to a function  
on $\cH$ by demanding that $\pounds_t \tilde v = 1$, see \eqref{com}.  Given $\tilde v$, we define a new initial cut $\tilde \cC$ by $\tilde v = 0$, and then we obtain a new GNC 
system $(\tilde v, \tilde r, \tilde x^A)$ from the foliation $\tilde \cC(\tilde v)$ with corresponding $\tilde k^a, \tilde s^a$ etc. 

At this point we have shown $\pounds_{\tilde k}( \tilde \alpha^{(0)}, \tilde \beta^{(0)}_a, \tilde \gamma^{(0)}_{ab} ) = 0$ on $\cH$ and $\tilde \alpha^{(0)} = \kappa^{(0)}$ on $\cH$. So we 
have all the equations in \eqref{ih} for $n=0$ and $m=0$ for the tilde fields. Now we wish to show the same for $m=1$ then $m=2$, and so on. To do this, we perform an induction in $m$. 
First, we consider the $\tilde v$-derivative of the $AB$ Ricci tensor component \eqref{ABR} which gives, on $\cH$
\ben
\pounds_{\tilde k} (R_{cd} p^c{}_a p^d{}_b)  = \pounds_{\tilde k}[\pounds_{\tilde k} \pounds_{\tilde l} \tilde \gamma_{ab} + \tilde \alpha \, \pounds_{\tilde l} \tilde \gamma_{ab}] + O(\ell^2).
\een
 Since the tensors in this equation are Lie-derived by $t^a = \tilde k^a + \tilde s^a$, 
and since the Ricci tensor is $O(\ell^2)$ by the Einstein equation, this gives, on $\tilde \cC$,
\ben
0= \pounds_{\tilde S}[\pounds_{\tilde S} \pounds_{\tilde l} \tilde \gamma^{(0)}_{ab} - \kappa^{(0)} \, \pounds_{\tilde l} \tilde \gamma^{(0)}_{AB}]
\een
see eq. (56) of \cite{hiw}. ``Integrating'' this equation along the orbits of $\tilde S^a$, it can be shown \cite{hiw} that $\pounds_{\tilde S} \pounds_{\tilde l} \tilde \gamma^{(0)}_{ab} = 0$, and then $ \pounds_{\tilde l} (\pounds_{\tilde k} \tilde \gamma^{(0)}_{ab})_{\tilde r=0} = 0$. This is the first equation of \eqref{ih} for $n=0$ and $m=1$. 

Let $M \ge 1$. We inductively assume that the first equation in \eqref{ih} is satisfied for $n=0$ and all $m \le M$ whereas the second and third equations in \eqref{ih} is satisfied for $n=0$ and all 
$m \le M-1$ (we have seen that this is true when $M=1$). Now we apply $\pounds_{\tilde l}^{M-1} \pounds^{}_{\tilde k}$ to the $\tilde v \tilde r$ component of the Ricci tensor \eqref{nlR} and evaluate the result at $\tilde r=0=r$, i.e. on $\cH$. Using the inductive hypothesis, we find 
\ben
\pounds_{\tilde l}^{M-1} \Big( \pounds^{}_{\tilde n} (R_{ab} \tilde l^a \tilde k^b) \Big)_{\tilde r=0} = \pounds_{\tilde l}^M \Big(\pounds_{\tilde k}^{} \tilde \alpha \Big)_{\tilde r=0} + O(\ell^2)
\een
and then using that the Ricci tensor is itself of order $O(\ell^2)$ by the Einstein equation, we 
obtain $ \pounds_{\tilde l}^M (\pounds_{\tilde k}^{} \tilde \alpha^{(0)})_{\tilde r=0} = 0$, which is the third equation in \eqref{ih} for $n=0$ and $m=M$.
Now we apply $\pounds_{\tilde l}^{M-1} \partial^{}_{\tilde v}$ to the $A \tilde r$ component of the Ricci tensor \eqref{lAR} and evaluate the result at $\tilde r=0$, i.e. on $\cH$.
We find 
\ben
\pounds_{\tilde l}^{M-1} \Big( \pounds^{}_{\tilde n} (R_{bc} \tilde p^b{}_a \tilde l^c)  \Big)_{\tilde r=0} = \pounds_{\tilde l}^M \Big(\pounds_{\tilde k}^{} \tilde \beta_a \Big)_{\tilde r=0} + O(\ell^2)
\een
which gives $ \pounds_{\tilde l}^M (\pounds_{\tilde k} \tilde \beta^{(0)}_a)_{\tilde r=0} = 0$. This is the second equation in \eqref{ih} for $n=0$ and $m=M$.
Thus we see that all equation of \eqref{ih} are satisfied for $n=0$ and all $m\ge 0$ hold for the tilde GNCs. 
For ease of notation, we finally replace the tilde cross section $\tilde \cC$ and the tilde tensor fields $\tilde \alpha, \tilde \beta_a, \tilde \gamma_{ab}, \tilde k^a, \tilde s^a, \tilde l^a$ so obtained by untilde quantities. 

\subsection{Induction step $n-2 \to n$, $(n\ge 2)$}

As the first step, we would like to show in parallel with the induction start that $\pounds_k \gamma_{ab} = O(\ell^{n+2})$ for $r=0$, i.e. on $\cH$. 
A first idea might be to consider again the $v$-derivative of the area functional, $\frac{\dd}{\dd v} A[\cC(v)]=0$, which vanishes due to stationarity. 
Again, let $\lambda$ be a parameter of affine null geodesics ruling $\cH$ with tangent $n^a$. In parallel with the induction start, it seems natural that we
 seek to combine the Raychaudhuri equation \eqref{Ray} and the Einstein equation \eqref{Ein:eq} as in 
\ben
\label{eq:Ray2}
\partial_\lambda \theta = -\sigma_{ab} \sigma^{ab} - \frac{1}{d-2} \theta^2 - \ell^2 H_{4 \, ab}n^an^b - \cdots -\ell^{2D-2} H_{2D \, ab}n^a n^b  
\een
in order to gain information on $\theta^{(n)}_{}, \sigma^{(n)}_{ab}$ at the induction order $n$. Indeed, the Raychaudhuri equation is implicitly used in the corresponding
argument for the induction start $n=0$ because it is the basis of area theorem used there \cite{he,wald}. In the present case,
the induction hypothesis gives $\theta^{(j)}_{}, \sigma^{(j)}_{ab}=0$ for all 
$j \le n-2$ on $\cH$, so $\theta, \sigma_{ab} = O(\ell^n)$. This means that the $\sigma_{ab} \sigma^{ab}, \theta^2$ terms in the Raychaudhuri equation \eqref{eq:Ray2} are of order $O(\ell^{2n})$. However, the other terms on the right side of \eqref{eq:Ray2} are potentially only of order $O(\ell^{n+2})$! This is because it is merely known at this stage that each term
$H_{j \,ab}n^a n^b$ is at least linear in positive boost weight quantities in the sense of \cite{hkr}, but when combined with lem. 2, the induction hypothesis, and the explicit $\ell^{j-2}$ powers, this 
still leaves room for a term only of order $O(\ell^{n+j-2})$ which could e.g. be as bad as $O(\ell^{n+2})$ (for $j=4$). So it appears that unlike for $n=0$, we cannot get useful 
sign information on $\partial_\lambda \theta^{(n)}$ from the Raychaudhuri equation as the sign-definite term is no longer leading in $\ell$ in the case $n \ge 2$ considered now. As a consequence, it is not easy to see how we could conclude $\theta^{(n)}_{}, \sigma^{(n)}_{ab} = 0$ [equivalent to the statement that $\gamma_{ab}$ is Lie derived by $k^a$ up to order $O(\ell^{n+2})$] on $\cH$ by some
sort of argument along the lines of the Raychaudhuri equation/area functional. Thus, it appears unclear how to take the first step in closing the induction. 

Below in lem. 3, we shall circumvent this problem by replacing the Raychaudhuri equation and cross section area by an equation for an entropy current-density and 
the corresponding generalized entropy of the cross section considered in \cite{hkr}. However, before we come to this construction, 
we observe that the Raychaudhuri equation still gives the following preliminary result.

\medskip
\noindent
{\bf Lemma 1.} Under the inductive hypothesis, we have $\theta^{(n)} = 0$ on $\cC$. 

\medskip
\noindent
{\em Proof:} Remember that combining the Einstein equation with the Raychaudhuri equation, we get \eqref{eq:Ray2}.
Now we expand $H_{j \, ab}n^an^b$ in terms of ``primitive monomials'' with definite ``boost weight'' as described in 
\cite{hkr}. By the results of sec. 2 of that paper, each summand in $H_{j \, ab}n^an^b$ is at least linear 
in a positive boost weight primitive monomial. By lem. 2 below, such a term is of order $O(\ell^{n})$, and since 
each $H_{j \, ab}n^an^b$ is accompanied by $\ell^{j-2}$ with $j-2 \ge 2$ we see that the corresponding terms 
on the right side of the above equation are of order at least $O(\ell^{n+2})$. By the inductive hypothesis 
$\theta, \sigma_{ab}$ are of order $O(\ell^{n})$, so since $n \ge 2$, the right side Raychaudhuri's equation
\eqref{eq:Ray2} is of order $O(\ell^{n+2})$. This shows that $\partial_\lambda \theta^{(n)} = 0$ on $\cH$. 

Now, since $n^a = (\pounds_k \lambda)^{-1} k^a$, and since $\pounds_t [(\pounds_k \lambda) \theta]=0$ because 
$[t,k]^a=0$, it follows that 
\ben
\pounds_k [(\pounds_k \lambda^{(0)}) \theta^{(n)}] = 
-\pounds_{S} [(\pounds_k \lambda^{(0)}) \theta^{(n)}]
\een
where $S^a = s^a|_{\ell=0}$.
By the usual relationship between affine- and Killing parameters at order $O(\ell^0)$, we can 
say that $\pounds_k \lambda^{(0)} = e^{\kappa^{(0)} v}$ (using the induction hypothesis and applying a rescaling to the affine parameter if necessary), which is constant 
on $\cC$. Since we have already seen that $\pounds_k \theta^{(n)}=0$, this gives 
\ben
(\pounds_{S} + \kappa^{(0)})\theta^{(n)} = 0
\een
on $\cC$. 
Let $\hat \phi_\tau$ be the flow of $S^a$ on $\cC$. The previous equation can be rewritten as 
\ben
\left(\frac{\dd}{\dd v} + \kappa^{(0)} \right)
\theta^{(n)}\circ \hat \phi_v  = 0
\een
Integrating this, we see that 
$
\theta^{(n)}\circ \hat \phi_v=e^{-\kappa^{(0)}v}X^{(n)}
$
where $X^{(n)}$ does not depend on $v$. 
However $\theta^{(n)}\circ \hat \phi_v$ is clearly bounded uniformly in $v$ because $\cC$ is compact, 
so letting $v \to -\infty$ and using that $\kappa^{(0)}>0$, we see that $X^{(n)}=0$. We therefore conclude that $\theta^{(n)}=0$ on $\cC$. \qed

\medskip
\noindent
{\bf Lemma 2}: Let $X$ be a primitive monomial as defined in \cite{hkr} of positive boost weight both defined relative to the GNC based on the 
affine parameter $\lambda$. Then we have $X = O(\ell^n)$ locally near $\cC$. 

\medskip
\noindent
{\em Proof.} 
Let us apply lem. 2.2 of \cite{hkr} to $X$. Then we eliminate any occurrence of the GNCs (with respect to the affine parameterization of $\cH$) 
of the Ricci tensor or its covariant derivatives using the Einstein equation \eqref{Ein:eq} and its covariant derivative. The new terms arising from the substitution process 
are decomposed into primitive monomials, and then lemma 2.2 is applied again, and the Einstein equation is used again etc., repeating this process $n/2$ times. 
Thereby, $X$ is written as a sum of terms which either have an explicit pre-factor of at least $\ell^{n}$, or 
terms which are products of the monomials described in lemma 2.2 of \cite{hkr} without the occurrence of the Ricci tensor. The terms of order $O(\ell^{n})$ 
can be ignored for the purposes of the proof, whereas each of the other terms contains at least one factor of $D_{c_1} \cdots D_{c_r} \pounds^N_n \sigma_{ab}$
or $D_{c_1} \cdots D_{c_r} \pounds^N_n \theta$
for some $N,r$. However, on $\cH$, we can replace $\pounds_n = (\pounds_k \lambda)^{-1} \pounds_k$, so such a term clearly is of order $O(\ell^{n})$ 
in view of the induction hypothesis. 
\qed

\medskip
\noindent
{\bf Lemma 3.} We have $\pounds_k \gamma_{ab} = O(\ell^{n+2})$ on $\cH$.

\medskip
\noindent
{\em Proof.}
In order to get around the problem described before lem. 1, we recall that the area is  the black hole entropy for Einstein gravity $\ell=0$, so it is natural to try an appropriate entropy 
functional for the higher derivative theory. We chose the ``improved IWW entropy'' $S[\cC]$ defined in\footnote{The difference between the ``improved IWW entropy'' and its antecedents
\cite{Bhat21,Wall2015} is a specific choice for the ambiguities in that construction which are designed to render $S[\cC]$ covariant, see prop. 1 of \cite{hkr}.} 
sec. 3.3 of \cite{hkr} because its properties are suitable for our purposes -- in particular it is covariant. 
$S[\cC]$ is defined in terms of an entropy-current-density $\sigma^a = \Sigma n^a + J^A (\partial_A)^a$ on $\cH$
as in 
\ben
S[\cC]= \int_{\cC} \Sigma \sqrt{\gamma} \dd^{d-2} x
\een
The following properties will be used, see prop. 1 of \cite{hkr}.
\begin{itemize}
\item
$\Sigma$ is a  local functional of boost weight 0 that for any cross section $\cC$ 
is a contraction of the following factors. Let $\bar n^a$ be tangent to a past directed congruence of affinely parameterized null geodesics 
transversal to $\cH$ normalized relative to $n^a$, as described in \cite{hkr}, and let 
\ben
\bar \theta = \gamma^{ab} \nabla_{(a} \bar n_{b)}, \quad \bar \sigma_{ab} = \gamma_a{}^c \gamma_b{}^d (\nabla_{(c} \bar n_{d)} - \tfrac{1}{d-2} g_{cd} \nabla_e \bar n^e)
\een
be the expansion and shear of that congruence. The possible factors are:
$$
 \cD_{(a_1} \cdots \cD_{a_j} \sigma_{a)b}^{}, \quad
 \cD_{(a_1} \cdots \cD_{a_j)} \theta, \quad
 \cD_{(a_1} \cdots \cD_{a_j} \bar \sigma_{a)b}^{}, \quad
 \cD_{(a_1} \cdots \cD_{a_j)} \bar \theta 
 $$
 where $\cD_a = D_a + \frac12 b \omega_a$ is a suitable covariantized derivative defined using the Hajicek 1-form $\omega_a$ \cite{hkr} and the boost weight $b$
it is acting on, or
 $$
 R^c{}_{a_1a_2d}, \quad \nabla_{(a_1} \cdots \nabla_{a_j} R^c{}_{a_{j+1} a_{j+2}) b} \quad (j>0)
 $$ 
 Products of these factor are contracted into 
 into $n^a$ or $\bar n^a$ or
$g^{ab}$ so as to yield a scalar, and 
boost weight 0 means that the total number of factors of the expansion/shear and of $\bar n^a$ is equal to the total number 
of factors of the expansion/shear and of $n^a$.

In particular, $S[\cC]$ is independent of the choices of affine parameter,
and hence a fully covariant functional of $\cC$ and the metric $g_{ab}$ in the sense that $S[\psi^{-1}(\cC),\psi^*g] = S[\cC,g]$ for any time-orientation preserving diffeomorphism $\psi$ preserving $\cH$. 

\item The analog of the Raychaudhuri equation which uses the Einstein equation \eqref{Ein:eq}:
\ben
\left( \pounds_n \nabla_a \sigma^a \equiv \right) \quad 
\partial_\lambda \left(
\frac{1}{\sqrt{\gamma}} \partial_\lambda (\sqrt{\gamma} \Sigma) + D_a J^a
\right) = F
\een
on $\cH$, where $F$ is at least quadratic in positive boost weight quantities, and where $J_a = J_A (\dd x^A)_a$ is an entropy current. [The quantity in parenthesis is just $\nabla_a \sigma^a$.]
\item $J^a$, which is a boost weight 1 quantity, contains an explicit power $\ell^2$  and is at least linear in positive boost weight terms. $\Sigma = 1 +$ zero boost weight terms 
containing an explicit power $\ell^2$.
\item $F = -\sigma_{ab}\sigma^{ab} - \frac{1}{d-2} \theta^2 + $ terms depending explicitly on $\ell^2$. 
\end{itemize}
\noindent
{\bf Example:} Consider the theory described by the action 
\ben
I[g] = \int_\cM (R + c_1 \ell^2 R_{ab} R^{ab} + c_2 \ell^2 R^2) \sqrt{-g} \dd^d x. 
\een
In this case, the improved IWW entropy density is \cite{Bhat19,Bhat21}
\ben
\Sigma = 1 + c_1 \ell^2(R_{ab}n^a \bar n^b - \theta \bar \theta) +2c_2 \ell^2 R.
\een
Note that the first term ``$1$'' corresponds to the area contribution to $S[\cC]$ so $S[\cC] = A[\cC] + O(\ell^2)$.
\medskip

Let us now return to the proof of lem. 3. By the first item, since $t^a$ is a KVF, since $S[\cC(v)]$ is a covariant functional that does not depend on the arbitrary choice
of affine parameter, and since the flow of $t^a$ by an amount $v$ moves $\cC$ to $\cC(v)$, we learn that 
$S[\cC(v)]$ is independent of $v$. Taking the first derivative in $v$ we get
\ben
\label{1st}
0= \frac{\dd}{\dd v} S[\cC(v)]_{v=0} = \int_{\cC} \frac{\partial \lambda}{\partial v} \, \partial_\lambda (\Sigma  \sqrt{\gamma}) \, \dd^{d-2}x . 
\een
Taking the second derivative in $v$ we get
\ben
\label{2nd}
0= \frac{\dd^2}{\dd v^2} S[\cC(v)]_{v=0} = \int_{\cC} \pounds_k \left[\frac{\partial \lambda}{\partial v} \, \partial_\lambda (\Sigma  \sqrt{\gamma})
\right] \, \dd^{d-2}x . 
\een
We will now use these two identities to prove the lemma.

First, since $k^a \nabla_a k^b = \alpha k^a$, and since $k^a = (\pounds_k\lambda) n^a$, we obtain $\alpha = \pounds_k \log \pounds_k \lambda$. 
Integrating this equation using $\alpha = \kappa + O(\ell^n)$ gives $\pounds_k\lambda = ae^{\kappa v}[1+O(\ell^n)]$ where $a(x^A)$ may be chosen to be $=1$ by a suitable choice of the affine parameter $\lambda$, giving 
\ben
\label{yyy}
D_a\pounds_k \lambda = O(\ell^n)
\een
on $\cC$.

Next, we look at $\sqrt{\gamma}^{-1} \partial_\lambda (\Sigma  \sqrt{\gamma})$. By the third item, we know that this is equal to $\theta +$ terms at least 
linear in positive boost weight with an explicit prefactor of at least $\ell^2$. Combined with lem. 1, lem. 2 and the induction hypothesis, we get that 
\ben
\label{xxx}
\frac{1}{\sqrt{\gamma}} \partial_\lambda (\Sigma  \sqrt{\gamma}) = O(\ell^{n+2})
\een
on $\cC$.

Finally, we use $\pounds_k \log \pounds_k \lambda
= \alpha$ which is used to write \eqref{2nd} as 
\ben
\label{3r}
\begin{split}
0 &= \int_{\cC} \alpha \, (\pounds_k \lambda) \,  \partial_\lambda (\Sigma  \sqrt{\gamma}) \, \dd^{d-2}x
+ \int_{\cC} (\pounds_k \lambda)^2 \partial_\lambda \left(
\frac{1}{\sqrt{\gamma}} \partial_\lambda (\sqrt{\gamma} \Sigma) + D_a J^a
\right)  \sqrt{\gamma} \, \dd^{d-2}x \\ 
&+ \int_{\cC} J^a D_a \pounds_k \lambda \,  \sqrt{\gamma} \, \dd^{d-2}x +
\int_{\cC} \theta \partial_\lambda (\sqrt{\gamma} \Sigma)
\, \dd^{d-2}x 
\end{split}
\een
where we have added and subtracted $D_aJ^a$ under the integral and performed a partial integration. The terms on the right side are now treated as follows. On the first term, we use that $\alpha$ is, by induction, constant on $\cC$ up to terms of order $O(\ell^n)$. The constant does not contribute in view of \eqref{1st}. The $O(\ell^n)$ contributes a term of order $O(\ell^{2n+2})$ in view of \eqref{xxx}. On the second term on the right side we 
use the analog of the Raychaudhuri equation in the second item. On the third term on the right hand side we use 
\eqref{yyy} and the fact that $\sigma^a$ is of order $O(\ell^{n+2})$ from the induction hypothesis and the third item. Thus, the third term is of order $O(\ell^{2n+2})$. 
The fourth term on the right side is treated using that $\theta=O(\ell^{n+2})$ by lem. 1 which is combined with \eqref{xxx}. Thus, the fourth term is of order $O(\ell^{2n+4})$. Combining these results, we 
see that \eqref{3r} gives us
\ben
\label{22}
\int_{\cC} (\pounds_k \lambda)^2 F \, \sqrt{\gamma} \dd^{d-2}x = O(\ell^{2n+2}). 
\een
At this stage, the fourth item gives together with the induction hypothesis that 
\ben
F = -\ell^{2n}
\sigma^{(n)}_{ab} \sigma^{(n)}_{cd} \gamma^{(0)ac} \gamma^{(0)bd} + O(\ell^{2n+2}), 
\een
and so 
\eqref{22} yields $\sigma^{(n)}_{ab} = 0$ on $\cC$
because $\pounds_k \lambda>0$. Note that we already know $\theta^{(n)} = 0$ by lem. 1, and that $\theta^{(j)} = 0 = \sigma_{ab}^{(j)}$ for $j \le n-2$ by induction.
Then, using the definitions of $\theta,\sigma_{ab}$ on $\cH$ we see that $\pounds_k \gamma_{ab} = O(\ell^{n+2})$ on $\cC$ and since $\pounds_t \pounds_k \gamma_{ab} = 0$ on $\cH$ it follows that $\pounds_k \gamma_{ab} = O(\ell^{n+2})$ everywhere on $\cH$. 
\qed

\medskip
Using $t^a = k^a + s^a$ and $\pounds_t \gamma_{ab} = 0$, we learn from lem. 3 that $\pounds_s \gamma_{ab} = O(\ell^{n+2})$. We now want to use this result to establish
the induction hypothesis \eqref{ihsn} at order $n$, i.e. that $s^{(j)a}$ are KVFs of the zeroth order metric $\gamma_{ab}|_{\ell=0}$ for all $j \le n$. This is in question only for $j=n$. It does not 
seem possible to deduce this merely from $\pounds_s \gamma_{ab} = O(\ell^{n+2})$ and the fact that $s^{(j)a}, j \le n-2$ are KVFs of the zeroth order metric. But we will now show that 
it can be achieved if we simultaneously redefine $\tilde t^a = \phi^* t^a, \tilde g_{ab} = \phi^*g_{ab}$ by a suitable diffeomorphism $\phi$ preserving the horizon cross sections $\cC(v)$. 
We will now construct such a diffeomorphism. The corresponding tensor fields $\tilde \alpha, \tilde \beta_a,  \tilde \gamma_{ab}, \tilde s^a, \tilde k^a$ will clearly still satisfy $\pounds_{\tilde k} \tilde \gamma_{ab} = O(\ell^{n+2}), \pounds_{\tilde k} \tilde \beta_{a} = O(\ell^n), \pounds_{\tilde k} \tilde \alpha = O(\ell^n)$. Furthermore, if $\phi$ is the identity up to order $O(\ell^2)$, then the zeroth order metric will be unchanged, $\gamma_{ab}|_{\ell=0} = \tilde 
\gamma_{ab}|_{\ell=0}$, and if $\tilde s^a \equiv \phi^* s^a = S^a + O(\ell^n)$ where $S^a := \sum_{j=0}^{n-2} \ell^j s^{(j)a}$, then $s^{(j)a} = \tilde s^{(j)a}, j \le n-2$, so the induction hypothesis \eqref{ihsn}
will still hold for $\tilde s^a$ up to order $j \le n-2$. Finally, if we even have $\phi^* s^a = S^a + \xi^a + O(\ell^{n+2})$ where $\xi^a$ is of order $O(\ell^n)$ and at the same time a KVF of the zeroth order metric $\gamma_{ab}|_{\ell=0}$, then all of the previous will still hold and in addition the induction hypothesis \eqref{ihsn} will hold for $\tilde s^a$ up to order $j \le n$.

We are going to construct $\phi = \phi_1$ as the flow $\phi_\tau$ at parameter value $\tau=1$ of a vector field $\zeta^a$ that is to be determined. First of all, the diffeomorphism should preserve the 
horizon cross sections $\cC(v)$, and this will be achieved choosing a $\zeta^a$ that is tangent to $\cC$ and such that $\pounds_k \zeta^a = 0$. In fact we will construct $\zeta^a$ initially on 
$\cC$ and then define it in a neighborhood of $\cH$ by the condition that $\pounds_k \zeta^a = 0 = \pounds_l \zeta^a$. Next, $\phi$ should be the identity up to order $O(\ell^2)$, and we will 
achieve this by choosing $\zeta^a = O(\ell^2)$. To analyze what requirements on $\zeta^a$ are imposed by the remaining conditions, we consider the Taylor series 
for $\phi^*_\tau s^a |_{\tau=1} = \tilde s^a$ with remainder around $\tau=0$:
\ben
\begin{split}
\tilde s^a =& \sum_{j=0}^M \frac{1}{j!} \frac{\dd^j}{\dd \tau^j} \phi^*_\tau s^a |_{\tau = 0} + \frac{1}{M!} \int_0^1 (1-\tau)^{M} \frac{\dd^{M+1}}{\dd \tau^{M+1}} \phi^*_\tau s^a \ \dd \tau\\
=& \sum_{j=0}^M \frac{1}{j!} \pounds_\zeta^j s^a + \frac{1}{M!} \int_0^1 (1-\tau)^{M}  \phi^*_\tau \pounds_\zeta^{M+1} s^a \ \dd \tau .
\end{split}
\een
Clearly, since $\zeta^a = O(\ell^2)$, the integral remainder term in the last line will be of order $O(\ell^{2(M+1)})$, so if we choose $2M \ge n$, then it will be of order $O(\ell^{n+2})$. 
Furthermore, if we knew that $\pounds_\zeta S^a = O(\ell^n)$, then it would automatically follow that each term in the sum for $j \ge 2$ would also be of order $O(\ell^{n+2})$, whereas 
the $j=1$ term can be written as $-\pounds_S \zeta^a$ up to order $O(\ell^{n+2})$. Thus, 
we would know that 
\ben
\tilde s^a = s^a  -\pounds_S \zeta^a + O(\ell^{n+2}).
\een
From this equation, we see that we would have $\ell^n \tilde s^{(n)a} = \ell^n s^{(n)a} - \pounds_S \zeta^a + O(\ell^{n+2})$. Since we would like to satisfy the 
the induction hypothesis \eqref{ihsn} at order $j \le n$, we must achieve that $\ell^n s^{(n)a} - \pounds_S \zeta^a$ is a KVF, called $\xi^a$, of order 
$O(\ell^n)$ of the zeroth order metric $\gamma_{ab}^{(0)}$ on $\cC$, up to an error term of size $O(\ell^{n+2})$. 
That a $\zeta^a$ with all these properties exists is established in the following lemma.

\medskip
\noindent
{\bf Lemma 4.} There exists a smooth vector field $\zeta^a=O(\ell^2)$ and a KVF $\xi^a = O(\ell^n)$ of $\gamma_{ab}^{(0)}$, both tangent to $\cC$,
such that $\ell^n s^{(n)a} = \xi^a + \pounds_S \zeta^a + O(\ell^{n+2})$, 
where $S^a = \sum_{k \le n-2} \ell^k s^{(k)a}$.

\begin{proof}
In this proof, quantities obtained by setting $\ell = 0$ are denoted by an overbar such as in $\bar \gamma_{ab}^{} = \gamma_{ab}^{(0)}$ etc. Consider the equation 
\ben
\label{weak}
\bar D^a \bar D_{(a} \zeta_{b)} = -\frac12
\bar D^a \gamma_{ab}, 
\een
where indices have been raised/lowered with the Riemannian metric $\bar \gamma_{ab}$ on $\cC$. The right side is 
$L^2$-orthogonal to the KVFs of $\bar \gamma_{ab}$, and since $\gamma_{ab} = \bar \gamma_{ab} + O(\ell^2)$, it 
is clearly of order $O(\ell^2)$. Now consider the usual weak formulation of this 
equation for $\zeta_a$ in the Sobolev space $W^{1,2}(\cC,T^*\cC)$ obtained by contracting it into a 
$\phi^a \in C^\infty(\cC,T\cC)$, integrating over $\cC$, and formally performing an integration by parts to move 
one derivative onto $\phi^a$. Then bilinear form so obtained from the left side of \eqref{weak} has a coercivity property expressed 
by the Poincar\'e type inequality
\ben
\| \bar D_{(a} X_{b)} \|_{L^2} \ge {\rm const.}
\| X_a - \bar P X_a \|_{L^2}
\een
where $\bar P$ is the $L^2$-projector onto the span of the KVFs of $\bar \gamma_{ab}$. By standard arguments, 
there is hence a weak, and a forteriori smooth, solution $\zeta^a$ which is unique modulo the addition of a 
KVF of $\bar \gamma_{ab}$. In particular, since the source in the equation for $\zeta^a$ is of order $O(\ell^2)$, 
we can choose $\zeta^a = O(\ell^2)$ itself. Using \eqref{weak}, we get
\ben
\begin{split}
\bar D^a \bar D_{(a}^{}(\pounds_S \zeta_{b)}^{} - \ell^n s^{(n)}_{b)}) = 
-\frac12 \bar D^a \pounds_s \gamma_{ab} + 
O(\ell^{n+2}) = O(\ell^{n+2}),
\end{split}
\een
using that 
$[\bar D_a, \pounds_S]=0$ (since $S^a$ is a 
KVF of $\bar \gamma_{ab}$ by the inductive assumption), and the fact that $s^a = s^{(0)a} + \ell^2 s^{(2)a} + \dots \ell^n s^{(n)a} + 
O(\ell^{n+2})$. The Poincar\'e inequality 
implies that the kernel of the operator 
$\bar D^a \bar D_{(a} X_{b)}$ consists 
precisely of the KVFs of $\bar \gamma_{ab}$.
It follows that $\bar \gamma^{ab}( \pounds_S \zeta_{b} - \ell^n s^{(n)}_b )$ is a KVF, which we call $\xi^a$, of $\bar \gamma_{ab}$ modulo $O(\ell^{n+2})$.
Thus, $\ell^n s^{(n)a} = \xi^a+ \pounds_S \zeta^a + O(\ell^{n+2})$, as desired.

We would finally like to show that $\xi^a = O(\ell^n)$. Let $\hat \phi_\tau$ be the flow of $S^a$ on $\cC$. Since $S^a$ is a KVF of $\bar \gamma_{ab}$, this flow is 
isometric and in particular area preserving. Therefore, by basic theorems in ergodic theory (see e.g. \cite{walters}), if $T_{a_1 \dots a_r}$ is a smooth tensor field on $\cC$, then the limit (orbit average)
\ben
\langle T_{a_1 \dots a_r} (x) \rangle := \lim_{T \to \infty} \frac{1}{T} \int_0^T \hat \phi^*_\tau T_{a_1 \dots a_r} (x) \ \dd \tau
\een
will exist in the sense of $L^p(\cC)$ for any $p \ge 1$ and any component in an orthonormal tetrad. Since $\hat \phi_\tau$ is isometric, so the same will hold for 
$\bar D_a$-derivatives of $T_{a_1 \dots a_r}$, so convergence even occurs in any Sobolev space $W^{p,q}(\cC), q \ge 0$, hence in the topology of any 
$C^\alpha(\cC)$ for any $\alpha \ge 0$, by an appropriate Sobolev embedding theorem. As a consequence, if $T_{a_1 \dots a_r} = O(\ell^M)$ for 
some $M$, then also $\langle T_{a_1 \dots a_r} \rangle = O(\ell^M)$. Furthermore, we have
\ben
\begin{split}
\langle \pounds_S T_{a_1 \dots a_r} (x) \rangle 
=& \lim_{T \to \infty} \frac{1}{T} \int_0^T \hat \phi^*_\tau \pounds_S T_{a_1 \dots a_r} (x) \ \dd \tau \\
=& \lim_{T \to \infty} \frac{1}{T}  \int_0^T \frac{\dd}{\dd \tau} \hat \phi^*_\tau T_{a_1 \dots a_r} (x) \ \dd \tau \\
=& \lim_{T \to \infty} \frac{1}{T} \hat \phi^*_\tau T_{a_1 \dots a_r} (x) \bigg|_0^T = 0, 
\end{split}
\een
where the limit exists pointwise and uniformly, together with that of all of its derivatives. Now we take the orbit average of the equation  $\ell^n s^{(n)a} = \xi^a+ \pounds_S \zeta^a + O(\ell^{n+2})$ and 
obtain $\langle \xi^a \rangle = O(\ell^n)$. 
Since $\xi^a$ is already known to be a KVF of $\bar \gamma_{ab}$ and since, by our genericity assumption 4) all such KVFs commute, it follows that $\hat \phi^*_\tau \xi^a = \xi^a$
so $\langle \xi^a \rangle = \xi^a$, and this gives $\xi^a = O(\ell^n).$
\end{proof}

We relabel the fields and coordinates thus obtained by the untilde ones. Then we know at this stage that $\pounds_k  \gamma_{ab} = O(\ell^{n+2}), \pounds_k  \beta_{a} = O(\ell^n), \pounds_k \alpha = O(\ell^n)$, and we know that \eqref{ihsn} holds up to and including order $j=n$.
We now proceed to the $vA$-component of the Einstein equation \eqref{Ein:eq} on $\cH$, which using the induction hypothesis \eqref{ih} and $\pounds_k \gamma_{ab} = O(\ell^{n+2})$ and the 
expression \eqref{nAR} for the corresponding Ricci-component reads
\ben
D_a \alpha - \frac{1}{2} \pounds_k \beta_a = \sum_{j = 4}^{2D} \ell^{j-2} H_{j \, cb}k^bp^c{}_a + O(\ell^{n+2}) \quad \text{on $\cH$.}
\een
By lem. 2 the right side is of order $O(\ell^{n+2})$. 
Using $t^a = k^a+s^a$, we therefore 
find 
\ben
D_a\alpha 
+ \frac12 \pounds_s \beta_a = 
O(\ell^{n+2})
\een
Following the same reasoning as in the induction start, we have to consider next a 
higher order version of \eqref{11}, i.e.  $-\pounds_s f+ \alpha f = \tilde \alpha$, demanding now that $\tilde \alpha \equiv \kappa  + O(\ell^{n+2})$, where $\kappa$ is the average of $\alpha$ over $\cC$ with respect to the volume element of $\gamma_{ab}$. Since we inductively know that $\alpha = \sum_{k \le n-2} \ell^k \kappa_{(k)} + O(\ell^{n})$, it follows that $\alpha-\kappa = O(\ell^n)$. 
By analogy with the induction start, we next seek to define a smooth function $f$ on $\cC$
such that
\ben
\pounds_s f - \alpha f = \kappa .
\een
Let $\hat \phi_\tau$ be the flow of $s^a$ on 
$\cC$. We define
\ben
\label{fdef}
f(x) = \kappa \int_0^\infty
{\rm exp} \left[
-\int_0^\sigma \alpha \circ \hat \phi_\tau(x) \dd \tau
\right] \dd \sigma.
\een
Using the relation $\alpha = \sum_{k \le n-2} \ell^k \kappa^{(k)} + O(\ell^{n})$ and $\kappa^{(0)} > 0$, we see that the 
$\dd \sigma$ integral converges absolutely 
for sufficiently small $|\ell|$, and furthermore, that $f = 1 + O(\ell^{n})$.
By analogy with the induction start, the new coordinate $\tilde v$ should now satisfy (see \eqref{2}) 
\ben
\pounds_s \tilde v = 1 + \frac{1}{\kappa} \left( \pounds_s \log f - \alpha \right) + O(\ell^{n+2}).
\een
a smooth solution for which exists by the following lemma. 

\medskip
\noindent
{\bf Lemma 5.} The equation $\pounds_s \varphi = \alpha - \kappa + O(\ell^{n+2})$ on $\cC$ has a smooth solution 
$\varphi = O(\ell^n)$.

\begin{proof}
In this proof, quantities obtained by setting $\ell = 0$ are denoted by an overbar such as in $\bar \gamma_{ab}^{} = \gamma_{ab}|_{\ell = 0}$ etc.
 We take $\varphi$ as the unique solution to 
\ben
\label{Felliptic}
\bar D^a \bar D_a \varphi = -\frac12 \bar D^a \beta_a
\een
that is $L^2$-orthogonal to the constant functions on $\cC$ which exists because the right side is $L^2$-orthogonal to 
the constant functions. By the usual elliptic regularity results, it follows 
that $\varphi(x,\ell)$ is jointly smooth in $(x,\ell)$. Let $S^a := \sum_{j \le n} \ell^j s^{(j)a}$. By \eqref{ihsn}, known at this stage up
to and including order $n$, we have $[\bar D_a, \pounds_S] = 0 = \pounds_S \bar \gamma_{ab}$, from which it follows that 
\ben
\begin{split}
\bar D^a \bar D_a (\pounds_S \varphi - \alpha)
=& \
\pounds_S \bar D^a \bar D_a  \varphi - \bar D^a \bar D_a \alpha \\
=& \
-\frac12 \pounds_S  \bar D^a \beta_a - \bar D^a \bar D_a \alpha \\
=& \
-\frac12 \bar D^a \pounds_S \beta_a - \bar D^a \bar D_a \alpha \\
=& \
O(\ell^{n+2}).
\end{split}
\een
In the last line we used $D_a\alpha 
+ \frac12 \pounds_s \beta_a = 
O(\ell^{n+2})$ and that $s^a = S^a + O(\ell^{n+2})$. It follows 
that $\pounds_s \varphi - \alpha$
is equal to a constant function on $\cC$
plus a function of order $O(\ell^{n+2})$.
Since $s^a$ is a KVF of $\gamma_{ab}$
modulo $O(\ell^{n+2})$, it follows that 
\ben
\begin{split}
\label{Fsplit}
\pounds_s \varphi - \alpha =& \
O(\ell^{n+2}) + \frac{1}{A[\cC]} \int_{\cC}
(\pounds_s \varphi - \alpha) \sqrt{\gamma} \dd^{d-2} x\\
=& \
O(\ell^{n+2}) - \frac{1}{A[\cC]} \int_{\cC}
\alpha \sqrt{\gamma} \dd^{d-2} x 
= O(\ell^{n+2}) - \kappa, 
\end{split}
\een
as desired. From the induction hypothesis, we know $\alpha-\kappa = O(\ell^n)$, so $\pounds_s \varphi = O(\ell^n)$. Using this and 
$\pounds_S \bar \gamma_{ab}=0$ and $s^a = S^a + O(\ell^{n+2})$ we now show that we modify $\varphi$ if necessary so that $\varphi = O(\ell^n)$ by analyzing 
the consequences of the relations
\ben
\label{Fcond}
s^{(0)a} D_a \varphi^{(j)} + \cdots +
s^{(j)a} D_a \varphi^{(0)}  = 0
\een
for ascending $j$ from $0$ to $n-2$: We first get $s^{(0)a} D_a \varphi^{(0)} = 0$. Since $s^{(0)a}$ is an irrational linear combination  
KVFs $\psi_1^a, \dots, \psi^a_N$ generating 
the $2\pi$-periodic flows of isometries of 
$\bar \gamma_{ab}$, it follows that 
$\psi_i^a D_a \varphi^{(0)} = 0$ for all $i=1, \dots, N$. As a consequence of the genericity 
assumption $s^{(j)a}$ is also a linear combination  
KVFs $\psi_1^a, \dots, \psi^a_N$ and therefore $s^{(j)a} D_a \varphi^{(0)} = 0$ for $j \le n$. Then we move on to $j=2$ and so on in \eqref{Fcond} and continue in a similar way establishing that $\psi_i^{a} D_a \varphi^{(j)} = 0$ for $j \le n-2$. Therefore, 
$\varphi-\sum_{j \le n-2} \ell^j \varphi^{(j)} = O(\ell^n)$ still fulfills \eqref{Fsplit}.
\end{proof}

Using the solution $\varphi$ given by lem. 5, we set
\ben
\label{vtildef}
\tilde v  := \frac{1}{\kappa} (\varphi+\log f )
\een
as a function on $\cC$. Then it follows that $\tilde v = O(\ell^n)$ on $\cC$ and then we extend $\tilde v$ to a function $\tilde v = v + O(\ell^n)$ 
on $\cH$ by demanding that $\pounds_t \tilde v = 1$, see \eqref{com}. Given $\tilde v$, we define a new initial cut $\tilde \cC$ by $\tilde v = 0$, and then we obtain a new GNC 
system $(\tilde v, \tilde r, \tilde x^A) = (v,r,x^A) + O(\ell^n)$ from the foliation $\tilde \cC(\tilde v)$ with corresponding 
$\tilde k^a$ and $\tilde s^a$. By construction, $\tilde \alpha$ is constant on $\tilde \cC$ modulo $O(\ell^{n+2})$, 
and then $\pounds_{\tilde k} \tilde \beta_a = 0$ 
modulo $O(\ell^{n+2})$. 

So at this point we have shown $\pounds_{\tilde k}( \tilde \alpha, \tilde \beta_a, \tilde \gamma_{ab} ) = O(\ell^{n+2})$ on $\cH$ and $\tilde \alpha = \kappa + O(\ell^{n+2})$ on $\cH$
and we can still assume the induction hypothesis \eqref{ih} for the tilde tensors up to and including order $n-2$ when $m\ge 1$. Furthermore, we have the induction hypothesis 
\eqref{ihsn} for $\tilde s^a$ on $\tilde \cC$. 

 Now we wish to show that we have $\pounds_{\tilde l}^m( \pounds_k( \tilde \alpha, \tilde \beta_a, \tilde \gamma_{ab} ) )_{\tilde r=0}= O(\ell^{n+2})$ for all $m$. To do this, we again perform an induction in $m$ similar to the induction start. First, we consider the $\tilde v$-derivative of the $AB$ Ricci tensor component \eqref{ABR} which gives 
 \ben
 \pounds_{\tilde k} (R_{cd} \tilde p^c{}_a \tilde p^d{}_d) = \pounds_{\tilde k}[\pounds_{\tilde k} \pounds_{\tilde l} \tilde \gamma_{ab} + \tilde \alpha \, \pounds_{\tilde l} \tilde \gamma_{ab}] + O(\ell^{n+2})
 \een
 on $\cH$.
Since the tensors are Lie-derived by $t^a = \tilde k^a + \tilde s^a$, we can effectively replace $\pounds_{\tilde k} = -\pounds_{\tilde s}$
and since the Ricci tensor is $O(\ell^{n+2})$ by the Einstein equation \eqref{Ein:eq} 
and the induction hypothesis ($\pounds_{\tilde k}$ produces at least one primitive factor 
appearing on the right side of the Einstein equation resulting in an $O(\ell^n)$ term by lem. 2, which is multiplied at least by $\ell^2$), this gives, on $\tilde \cC$,
\ben
\pounds_{\tilde S}[\pounds_{\tilde S}  \tilde L_{ab} - \kappa \, \tilde L_{ab}] = O(\ell^{n+2})
\een
where $\tilde S^a = \sum_{j \le n} \ell^j \tilde s^{(j)a}$ which is a Riemannian isometry of $\tilde \gamma_{ab}|_{\ell = 0}$ by \eqref{ihsn}, and where 
$\tilde L_{ab}:=\pounds_{\tilde l} \tilde \gamma_{ab}$. ``Integrating'' this equation along the orbits of $\tilde S^a$ gives 
$\pounds_{\tilde S}  \tilde L_{ab}= O(\ell^{n+2})$ by the same kind of argument as around eq. (56) of \cite{hiw} but carrying around now 
the potential $O(\ell^{n+2})$ ``error terms''. This results in  
$ \pounds_{\tilde l} (\pounds_{\tilde k} \tilde \gamma_{ab})_{\tilde r=0} = O(\ell^{n+2})$, 
where we have replaced again $\pounds_{\tilde k} = -\pounds_{\tilde s}$, so we get the first equation of \eqref{ih} for our induction order $n$ and $m=1$. 

Let $M \ge 1$. We inductively assume that the first equation in \eqref{ih} is satisfied for all $m \le M$ whereas the second and third equations in \eqref{ih} is satisfied for all 
$m \le M-1$ (we have seen that this is true when $M=1$). Now we apply $\pounds_{\tilde l}^{M-1} \pounds^{}_{\tilde k}$ to the $\tilde v \tilde r$ component of the Ricci tensor 
\eqref{nlR} and evaluate the result at $\tilde r=0$, i.e. on $\cH$. Using the inductive hypothesis and the Einstein equation, we find 
\ben
\pounds_{\tilde l}^{M-1} \Big( \pounds_{\tilde k} (R_{ab} \tilde l^a \tilde k^b) \Big)_{\tilde r=0} = \pounds_{\tilde l}^M \Big(\pounds_{\tilde k}^{} \tilde \alpha \Big)_{\tilde r=0} + O(\ell^{n+2})
\een
and then using that a $\tilde v$-derivative of the Ricci tensor is itself of order $O(\ell^{n+2})$ by the Einstein equation ($\pounds_{\tilde k}$ 
hits at least one primitive factor of non-negative boost weight
appearing on the right side of the Einstein equation resulting in an $O(\ell^n)$ term which is multiplied at least by $\ell^2$), we 
obtain $ \pounds_{\tilde l}^M (\pounds_{\tilde k}^{} \tilde \alpha)_{\tilde r=0} = O(\ell^{n+2})$, which is the third equation in \eqref{ih} for our induction order $n$ and $m=M$.
Now we apply $\pounds_{\tilde l}^{M-1} \pounds^{}_{\tilde k}$ to the $A \tilde r$ component of the Ricci tensor \eqref{lAR} and evaluate the result at $\tilde r=0$, i.e. on $\cH$.
We find 
\ben
\pounds_{\tilde l}^{M-1} \Big( \pounds_{\tilde k} (R_{bc} \tilde l^c \tilde p^b{}_a) \Big)_{\tilde r=0} = \pounds_{\tilde l}^M \Big(\pounds_{\tilde k}^{} \tilde \beta_a \Big)_{\tilde r=0} + O(\ell^{n+2})
\een
which similarly gives $ \pounds_{\tilde l}^M (\pounds_{\tilde k} \tilde \beta_a)_{\tilde r=0} = O(\ell^{n+2})$. This is the second equation in \eqref{ih} for $m=M$.
Thus we see that all equations of \eqref{ih} are satisfied for  all $m\ge 0$ for the tilde GNCs. 
We relabel the tilde GNCs and corresponding tensors by the untilde $(v,r,x^A)$ to simplify the notation. This closes the induction loop.

\medskip

Setting $\chi^a := k^a$ where $k^a$ is the vector field that is defined in a neighborhood of $\cH$ 
by going through $n$ iterations of the induction step as described above,  
we obtain the following theorem.

\medskip
\noindent
{\bf Theorem 1.}
Suppose that we have a family of spacetimes satisfying the assumptions 1)-4) in sec. 2.1 and let $n \in {\mathbb N}_0$. Then there exists a vector field $\chi^a$ tangent to the null generators of $\cH$ 
which Lie derives $g_{ab}$ modulo terms of order $O(\ell^{n+2})$ and modulo terms that vanish to arbitrarily high order in any coordinate transverse to $\cH$. 
In other words, if $l^a$ is a VF transverse to $\cH$ (such as $l^a$ in the GNC above), then 
\ben
\pounds_l^m \pounds_\chi g_{ab} |_\cH = O(\ell^{n+2})
\een
for any $m \ge 0$.
$\chi^a$ commutes with $t^a$ and on 
$\cH$ satisfies $\chi^a \nabla_a \chi^b = \kappa \chi^b$, where $\kappa$ is constant up to terms of order $O(\ell^{n+2})$. 
The vector field $s^a = t^a-\chi^a$ is tangent to a foliation of cross sections $\cC(v)$ of $\cH$ and on $\cH$ satisfies
\ben
\pounds_l^m \pounds_s g_{ab} |_\cH = O(\ell^{n+2}).
\een

\section{Non-rotating case}
\label{sec:nonrot}

 Now we assume at first that we are in the non-rotating case II): $t^a$  is tangent to the null generators of $\cH$ for sufficiently small $|\ell|$. We will show order by order in $\ell$ that $(\cM,g_{ab})$ is spherically symmetric. 

First, for $\ell=0$ it follows from the staticity theorem \cite{sudarsky} in combination with \cite{CW94} and the uniqueness theorems for static vacuum black holes in Einstein gravity \cite{Israel67,Israel68,bunting,gibbons,gibbons1,Gibbons02a,ruback,Rogatko:2003kj} that the metric $\bar g_{ab}:=
g_{ab}|_{\ell=0}$ is the Schwarzschild metric i.e. there is a coordinate system in which 
\ben
\bar g = -f \dd t^2 + f^{-1} \dd r^2 + r^2 \dd \Omega^2_{d-2}, \quad f=1-\left( \frac{r_0}{r} \right)^{d-3}, \quad r_0 > 0
\een
with $\dd \Omega^2_{d-2}$ the metric of the round sphere ${\mathbb S}^{d-2}$.
By applying a suitable $\ell$-dependent diffeomorphism to the family $g_{ab}(\ell, x)$ which is the identity to zeroth order in $\ell$, we 
can ensure that the timelike KVF $t^a$ is independent of $\ell$. We assume that such a diffeomorphism has been applied and continue to call the family of metrics $g_{ab}(\ell, x)$.
In particular, $\bar g_{ab}$ is still given by the above formula. By assumption  $t^a$ is null on $\cH$.
 
Let $Y^a$ be one of the KVFs of the Schwarzschild metric generating a rotation. We will now construct a formal series 
$Y^a(\ell, x) = \sum_{k \ge 0} \ell^k Y^{(k)a}(x)$ in $\ell$ which is tangent to $\cH^\pm$, which is commuting with $t^a$, which Lie-derives $g_{ab}(\ell, x)$ to all orders in $\ell$, 
and such that  $Y^a(\ell=0,x) = Y^a(x)$. For this, we assume inductively that the terms in this expansion for $Y^a$ have been constructed up to and including order $O(\ell^{n-2})$ in such a way that 
$\pounds_Y g_{ab}=O(\ell^n)$ and $\pounds_t Y^a = O(\ell^n)$. If we now take $\pounds_Y$ of the Einstein equation, evaluate this at order $\ell^n$, then we see that 
$h_{ab}$, defined as the $O(\ell^n)$-term in $\pounds_Y g_{ab}$, satisfies the homogeneous linearized Einstein equation in Schwarzschild. Furthermore, since the linearized Einstein operator of Schwarzschild 
commutes with $\pounds_t$, since $\pounds_t Y^a = O(\ell^n)$ and since $\pounds_t g_{ab} = 0$, we have $\pounds_t h_{ab}=0$. Thus, $h_{ab}$ is a stationary perturbation of Schwarzschild which is asymptotically flat, i.e. falling off roughly 
as $|h_{\mu\nu}| = O(1/r^{d-3})$ as $r\to \infty$ in a suitable asymptotically Cartesian coordinate system $(x^\mu)$, and regular on the horizon, see def. 2.1 of \cite{CW94}
for the details on such asymptotic conditions. 

\medskip
\noindent
{\bf Proposition 1.}
Let $h_{ab}$ be a linearized, smooth, asymptotically flat solution to the linearized Einstein equation off of Schwarzschild spacetime which is regular on $\cH$ and Lie-derived by $t^a$. Then 
\ben
h_{ab} = z_{ab} + \pounds_\xi \bar g_{ab}
\een
where $\xi^a$ is a smooth vector field which is an asymptotic symmetry at null infinity and where $z_{ab}$
is a perturbation towards a Myers-Perry black hole \cite{mp}.

\begin{proof}
This proof is based on the analysis of master variables for gravitational perturbations given in \cite{ishibashi,ishibashi1}, see app. \ref{app:B}  for the full argument. 
\end{proof}

We learn from prop. 1 that 
\ben
\pounds_Y g_{ab} = \ell^n(z_{ab} + \pounds_\xi \bar g_{ab}) + O(\ell^{n+2})
\een
where $z_{ab}$ is an infinitesimal perturbation to a Myers-Perry black hole and where $\xi^a$ is a gauge vector field which is smooth as $r \to r_0$, and which is an asymptotic symmetry at $\cI^+$ and $\cI^-$, i.e. has one of the asymptotic forms IIa,b,c, III, IV of \cite{hiw}. From $\pounds_t h_{ab}=0=\pounds_t z_{ab}$ it also follows that $\pounds_{[t,\xi]} \bar g_{ab} = 0$ so $\eta^a:=[t,\xi]^a$ must be a linear combination of $t^a$ and a rotational KVF of Schwarzschild. By inspection, the only  asymptotic forms for $\xi
^a$ giving rise to a non-trivial asymptotic symmetry $\eta
^a$ are type III, i.e. asymptotic boosts, in which case $\eta^a$ is an asymptotic spatial translation. This, however, is not a KVF of Schwarzschild, so we conclude that $[t,\xi]^a=0$, in fact. It follows that $\bar g_{ab} t^a \xi^b$ is Lie-derived by $t^a$ on $\cH$, and since the latter vanishes on the bifurcation surface $\cB$, we must have $\bar g_{ab} t^a \xi^b = 0$ on $\cH$
meaning that $\xi^a$ is tangent to $\cH$ because $t^a$
is null with respect to $\bar g_{ab}$.

Now we redefine $Y^a \to Y^a + \ell^n \xi^a$ which is tangent to $\cH$. Then still $[t,Y]^a = 0$, 
$\pounds_Y g_{ab}=O(\ell^n)$ but now $\pounds_Y g_{ab}=\ell^n z_{ab}+O(\ell^{n+2})$. Furthermore, since by assumption $s_c:=g_{ab} t^a p^b{}_c|_{\cH} = 0$ to all orders in 
$\ell$ by the non-rotating assumption, since $t^a$ does not depend on $\ell$, and since 
$\pounds_Y t^a = 0$, we get
$
N_c := z_{ab} t^a p^b{}_c = 0
$
on $\cH$, so $z_{ab}$ cannot be an infinitesimal 
perturbation towards a rotating black hole, for which $N_c$ would be the perturbed shift vector on $\cH$ which is not zero. 
Thus, $z_{ab}$ must be a perturbation towards another Schwarzschild black hole. Consider now the pull-back of the equation $\pounds_Y g_{ab}=\ell^n z_{ab} + O(\ell^{n+2})$  to the bifurcation surface $\cB$. Since $Y^a$ is tangent to 
$\cB$, we get $D_a Y_b + D_b Y_a= \ell^n z_{ab} + O(\ell^{n+2})$
on $\cB$. Taking a trace of these equation and integrating over $\cB$ shows (in our gauge where the location of $\cB$
is independent of $\ell$)
\ben
0=2 \int_{\cB} D_a Y^a \sqrt{\gamma} \dd^{d-2} x  =
\ell^n \int_{\cB} z_{ab}^{} \gamma^{ab} \sqrt{\gamma} \dd^{d-2} x + O(\ell^{n+2})
\een
and this implies that $z_{ab} = 0$ because a non-trivial perturbation to another Schwarzschild black hole will result in a change of the area of $\cB$.
This closes the induction loop, showing the existence of a KVF $Y^a(\ell, x)$ commuting with $t^a$, to all orders in $\ell$. 

In the above argument we can start with any rotational KVF of Schwarzschild, so we obtain from the above construction not only one, but in fact $\tfrac{1}{2}(d-2)(d-1)$ KVFs
$Y^a_{j}(\ell, x), j=1, \dots, \tfrac{1}{2}(d-2)(d-1)$
commuting with $t^a$, to all orders in $\ell$. Generalizing the usual argument, see e.g.
\cite{wald}, app. C, that the space of KVFs on a pseudo-Riemannian manifold is finite-dimensional to formal series of VFs and metrics (in $\ell^2$) we learn that the $Y^a_{j}(\ell, x)$ generate a finite dimensional Lie-algebra under the commutator of VFs. This Lie algebra must be a deformation/extension of the Lie algebra $\frak{so}(d-1)$, and such extensions are classified by the cohomology ring
$H^2(\frak{so}(d-1), V)$ where $V$ is a finite-dimensional representation of $\frak{so}(d-1)$, see e.g. \cite{varadarajan}. As is well-known, that ring is trivial for any simple Lie-algebra and finite-dimensional representation. Thus the $Y^a_{j}(\ell, x)$ generate 
the Lie-algebra $\frak{so}(d-1)$.

We therefore have shown:

\medskip
\noindent
{\bf Theorem 2.}
In the non-rotating case, $g_{ab}(x,\ell=0)$ is a 
Schwarzschild metric. If $Y^a_{}(x)$ is one of its rotational KVFs, there is a formal series $Y^a_{}(\ell, x) = \sum_{k \ge 0} \ell^k Y^{(k)a}(x)$ in $\ell$ which is tangent to $\cH^\pm$, which is commuting with $t^a$, which Lie-derives $g_{ab}(\ell, x)$ to all orders in $\ell$, and such that  $Y^a_{}(\ell=0,x) = Y^a_{}(x)$.
The Killing vector fields $Y^a_{}(\ell, x)$ 
(in the sense of formal series)
represent the Lie algebra of $SO(d-1)$ under the vector field commutator.

\section{Existence of global rotational Killing field(s)} 
\label{sec:analytic}

Consider the rotating case I) and assume in addition to 1)--4) in sec. 2.1 that the manifold $\cM$ is real analytic and the metric $g_{ab}(\ell, x)$ and stationary KVF $t^a(\ell, x)$ are jointly real analytic in $(\ell, x)$ for some atlas of analytic coordinate systems (depending possibly on $\ell$). Let us go through the proof of thm. 1 with an eye towards analyticity of $\chi^a$
in $x$ at the various orders in $\ell$. First, we may pick the initial cut $\cC$ to be an analytic submanifold of the analytic manifold $\cH$. This means that 
initial GNCs $(v,r,x^A)$ give analytic charts in neighborhoods of $\cM$ covering $\cH$. As we have described, the vector field $\chi^a = \tilde k^a = (\partial/\partial \tilde v)^a$ described in 
thm. 1 is constructed as a coordinate vector field for a suitable new GNC system $(\tilde v, \tilde r, \tilde x^A)$, and this coordinate system is analytic as we will now argue. 
At $n$-th order in $\ell^n$, the function $\tilde v$ is defined by \eqref{vtildef} in terms of functions $f,\varphi$. The function $f$ is defined by \eqref{fdef}, and easily checked to be 
analytic \cite{hiw}. The function $\varphi$ is defined to be a solution to an elliptic equation \eqref{Felliptic} with analytic coefficients on $\cC$, hence also analytic 
by standard results on elliptic regularity \cite{hormander}. Thus $\tilde v$ is analytic. Likewise, the coordinates $\tilde x^A$ are constructed using the vector field $\zeta^A$ in lem. 4, 
and this vector field is analytic because it is also defined as the solution to an elliptic equation with analytic coefficients. The coordinates $(\tilde v, \tilde x^A)$ are propagated by Lie-transport 
with the analytic vector field $t^a$, and so are analytic functions on the respective coordinate patches of $\cH$.  Finally, $\tilde r$ is defined as a parameter along affine geodesics 
off of $\cC$ with analytic initial condition, hence it is also analytic. Thus, the coordinate systems $(\tilde v, \tilde r, \tilde x^A)$ successively determined at the various orders in $\ell$ 
are all analytic. Since $\chi^a$ is a coordinate vector field in this coordinate system, it is analytic, and we have $\pounds_\chi g_{ab} = O(\ell^{n+2})$ identically in 
an open neighborhood of $\cH$ for the given $n$ that we fix, by thm. 1.

At this stage, we can extend $\chi^a$ globally on the domain of outer communication 
\ben
\cD := I^+ \left(\bigcup_{t \in {\mathbb R}} \phi_t[\Sigma_1] \right) \cap  I^- \left(\bigcup_{t \in {\mathbb R}} \phi_t[\Sigma_1] \right)
\een
of $\cM$ (where $\Sigma_1$ is the asymptotic region of the acausal surface and $\phi_\tau$ the flow of $t^a$, see \cite{CW94} def. 2.1 and footnote \ref{foot:1})
by the usual method of analytic extension on overlapping neighborhoods. 
Because that domain is simply connected $\pi_1(\cD) = 0$ by the topological censorship theorem \cite{fsw,Galloway99} the analytic continuation is single-valued. Then proceeding 
precisely in the same  way as in sec. 3 of \cite{hiw}, which does not use the Einstein equations, we get in view of thm. 1:

\medskip
\noindent
{\bf Theorem 3.} Assume 1)--4) as in sec. 2.1, that $\cM$ is  real analytic and that $g_{ab} = g_{ab}(\ell, x), t^a = t^a(\ell,x)$ are jointly real analytic 
in $(\ell, x)$ with $t^a$ not tangent to the null generators of $\cH$. Then for any fixed $n \in {\mathbb N}_0$, there exist an analytic VF $s^a = s^a(\ell,x)$ on $\cD$ tangent 
to a foliation of $\cH$ by spacelike cross sections, and a VF $\chi^a=\chi^a(\ell,x)$ on $\cD$ tangent and normal to $\cH$, such that  
$s^a, \chi^a$ commute with $t^a$, such that $\pounds_s g_{ab} = O(\ell^{n+2}) = \pounds_\chi g_{ab} $ and such that we have
\ben
t^a = \chi^a + s^a
\een
The vector field $\chi^a$ has an acceleration (surface gravity) $\kappa$ on $\cH$ that is constant up to order $O(\ell^{n+2})$
 and $s^a$ can be written as
 \ben
 s^a = \sum_{j=1}^N \Omega_j \psi_j^a,
 \een
where $\psi_j^a=\psi_j^a(\ell,x)$ are commuting VFs on $\cD$ whose orbits are closed with period $2\pi$ 
such that $\pounds_{\psi_j} g_{ab} = O(\ell^{n+2})$, and 
where the $\Omega_j = \Omega_j(\ell)$ are constants that are defined up to order $O(\ell^{n+2})$. 

\medskip
\noindent
{\bf Remarks.} 1) A similar argument will work with a cosmological constant $\Lambda<0$, because the asymptotic structure is used in our proofs only 
to show that the domain of outer communication is simply connected and to show that $t^a$ does not vanish on $\cH$. Both 
will work if $(\cM,g_{ab})$ is asymptotically AdS. In the case $\Lambda>0$, one has to make suitable assumptions on how $t^a$ behaves on $\cI^-$ because this is spacelike now.
To stay within the realm of effective field theory, the cosmological constant should in either case be so small that $|\Lambda L^2| \lesssim 1$, where $L$ is the typical scale over which 
the solution is varying, as explained in \cite{hkr}, def. 2.1.

\noindent
2) Note that the VFs in the theorem $s^a, \psi_j^a, \chi^a$ in principle depend on the order $n$ in $\ell$ up to which the Killing vector field property holds. 
It should also be possible to establish the existence of $s^a, \psi_j^a, \chi^a$ with the properties stated in this theorem up to arbitrary order in $\ell$. 
This would be tantamount to showing that the successive changes of coordinates to $(\tilde v, \tilde r, \tilde x^A)$ defined order by order in $\ell$
can be summed within a non-zero radius of convergence $|\ell|<\ell_0$. We expect that this should be possible, but it would be a rather tedious bookkeeping 
exercise. We will not carry this out here because in the effective field theory spirit, the action $I[g]$ is only valid up to a finite order in $\ell$ anyhow.  

\noindent
3) The proof of thm. 1 shows that, by applying a suitable $\ell$-dependent
diffeomorphism to $g_{ab}, \chi^a$ and $t^a$, we can make the rotational KVFs $\psi_i^a, i=1, \dots, N$ independent of $\ell$.

\medskip

Consider next the non-rotating case II). By thm. 2, we 
have a set of KVFs in the sense of formal power series generating the Lie algebra of $SO(d-1)$ under the commutator of VFs. These KVFs are constructed order by order in $\ell^2$, and if we could show that the series converges, then it would follow that we have actual KVFs and not just formal series. Again, for such an argument to proceed we should at least know that the manifold $\cM$ is real analytic and that the metric $g_{ab}(\ell, x)$ is jointly real analytic 
in $(\ell, x)$ for an analytic atlas of coordinate systems depending analytically on $\ell$. Again, one would have to go through the detailed steps of the inductive constructions, order by order in $\ell^2$.

\section{Conclusions}
For simplicity, we have considered in this paper (parity even) purely gravitational theories. However, we expect our proofs to be robust and to apply to any local covariant Einstein-gravity-matter model such that the rigidity theorem holds for the corresponding standard Einstein-gravity-matter model when $\ell = 0$. The latter applies to a broad class of models including abelian vectors coupled to scalars \cite{hi,hiw}.  

A more difficult question is whether one can remove the analyticity, non-degenerate, and genericity assumptions in our main theorem (thm. 3). Since it is unknown how to remove non-degeneracy and analyticity even for Einstein gravity, we expect this to be highly non-trivial. Actually, in the case of higher derivative theories as considered in this paper, it is not totally clear what viewpoint to take on this problem, for the following reason. We have treated solutions in the EFT setting as (locally) small  corrections to the corresponding solutions in Einstein gravity. In doing so, we are assuming in effect that the solutions are ``low frequency' (relative to the EFT length scale $\ell$). If we were to try to drop the analyticity assumption, it seems plausible that we would have to study the EFT equations as an initial value problem of some sort as in \cite{klainerman}, which requires the study of solutions of arbitrary frequency. 
However, for a general higher derivative theory, there is at this point no general understanding when it will possess a well-posed initial value problem, although this has been 
established for certain special theories \cite{noakes, harvey1, harvey2}.

Let us assume that we {\em have} a higher derivative theory with a well-posed initial value problem, say one of the theories with second order equations of motion studied in \cite{harvey1, harvey2}. If we take such a theory seriously even for arbitrarily short wavelengths, it is not natural to consider the lightcone as defined by the metric $g_{ab}$ -- as we have done in this paper -- but instead one wants to study an intrinsically defined propagation cone defined by the highest derivative part in the Einstein equation. This leads in general to a propagation cone different from the lightcone and correspondingly a different notion of event horizon of a black hole \cite{Reall:2021voz}. One might ask in such special theories whether the rigidity theorem still holds for the new notion of event horizon, and whether, in the case of stationary solutions, the different notions of event horizons may actually even coincide. We leave this interesting issue for future work.

\medskip
\noindent
{\bf Acknowledgements:} S.H. thanks the Max-Planck Society for supporting the collaboration between MPI-MiS and Leipzig U., Grant Proj. Bez. M.FE.A.MATN0003.
The work of A.I. was supported in part by JSPS KAKENHI Grants No. 21H05182, 21H05186, 20K03938, 20K03975, 17K05451,
and 15K05092. H.S.R. is supported by STFC grant no. ST/T000694/1.

\appendix

\section{Ricci tensor in GNCs \cite{hiw}}\label{sect:A}

In this Appendix, we provide expressions for the Ricci tensor in our GNC system for the convenience of the reader \cite{hiw}. 
As in the main text, the horizon, $\cH$, corresponds to the surface $r=0$. Associated 
with the foliation $\cC(v,r)$ by $(d-2)$-dimensional compact cross sections there are two natural projectors. The orthogonal projector $q^a{}_b$, as well
as the non-orthogonal projector $p^a{}_b$ characterized by $p^a{}_b k^b = p^a{}_b l^b = 0$. When $r \beta_a \neq 0$, these do not coincide. 
In terms of the Gaussian null coordinate components 
of $\gamma_{ab}$, we have
$q^{ab} =
(\gamma^{-1})^{AB} (\partial x^A)^a (\partial/\partial x^B)^b$, 
whereas $p^a{}_b = (\partial/\partial x^A)^a (\dd x^A)_b$. 
The relationship between ${p^a}_b$ and ${\gamma^a}_b$ is given by
\ben
{p^a}_b = -r l^a \beta_b + {\gamma^a}_b \,.  
\een
Since in terms of Gaussian null coordinates, we have
$p^a{}_b = (\partial_A)^a (\dd x^A)_b$ and $l^a = (\partial_r)^a, k^a = (\partial_v)^a$, and since coordinate
vector fields commute, it follows that
$\pounds_k p^a{}_b = 0 = \pounds_l p^a{}_b$. It also is easily seen that
$q^{ac} \gamma_{cb} = p^a{}_b$ and that $p^a{}_b q^b{}_c = q^a{}_c$.
We finally recall that a definition $D_a$ of an intrinsic derivative operator associated 
with $q_{ab}$ was given in \eqref{Ddef}, and 
we denote the Riemann and Ricci tensors associated with $q_{ab}$ as
${R}[\gamma]_{abc}{}^d$ and ${R}[\gamma]_{ab}$. Then we have:
\bena 
\label{nnR} 
 k^ak^bR_{ab} &=&
 -\half q^{ab} \pounds_k\pounds_k \gamma_{ab} 
 +\quater q^{ca}q^{db}(\pounds_k \gamma_{ab})\pounds_k \gamma_{cd} 
 +\frac{1}{2} \alpha \: q^{ab}\pounds_k \gamma_{ab} 
\non \\
 &+&
 \frac{r}{2} \cdot 
 \Bigg[\;     
       4 \alpha \pounds_l \pounds_l \alpha 
       + 8 \alpha \pounds_l\alpha 
       + (\pounds_l\alpha) q^{ab}\pounds_k\gamma_{ab} 
\non \\ 
 && \qquad \,
      + q^{ab}\pounds_l\gamma_{ab} \cdot 
                \Big\{
                      -\pounds_k \alpha
                      -r q^{cd}\beta_c\pounds_k\beta_d 
\non \\
 && \qquad \qquad \qquad \qquad \,\,\: 
                      +(rq^{cd}\beta_c\beta_d+2\alpha)\pounds_l(r\alpha)
                      + rq^{cd}\beta_c D_d \alpha 
                \Big\}
\non \\ 
 && \qquad \, 
     + 2q^{ab}{D}_a  
          \left\{
             \beta_b\pounds_l(r\alpha) + {D}_b\alpha-\pounds_k\beta_b
          \right\} 
\non \\ 
 && \qquad \, 
     +q^{bc}\pounds_l(r\beta_c)\cdot
          \Big\{
                 (rq^{ef}\beta_e\beta_f+2\alpha) \pounds_l(r\beta_b) 
\non \\ 
 && \qquad \qquad \qquad \qquad \quad 
                - 4 D_b\alpha  
                + 2 \pounds_k\beta_b + 4rq^{ae}\beta_e {D}_{[a}\beta_{b]} 
          \Big\}
\non \\ && \qquad \,
     + 2(\pounds_l\alpha)\pounds_l(r^2q^{ab}\beta_a\beta_b) 
     + 4rq^{ab}\beta_a\beta_b \pounds_l\alpha 
     + 2 rq^{ab}\beta_a\beta_b \pounds_l \pounds_l \alpha 
\non \\ && \qquad \,  
     +2 q^{ab}\beta_a\pounds_l(r\beta_b)\cdot
      \left\{
             2 \pounds_l(r\alpha)
            - \half r q^{cd}\beta_c\pounds_l(r\beta_d)
      \right\} 
\non \\ 
 && \qquad \, 
     + 2r^{-1}\pounds_l \left\{ 
                                 r^2q^{ab}\beta_a(D_b\alpha-\pounds_k\beta_b)
                           \right\} 
     + 2r^{-1}\alpha \pounds_l(r^2 q^{ab}\beta_a\beta_b) 
 \Bigg] \,,
\eena

\bena 
\label{nlR} 
 k^al^bR_{ab}
     &=& -2 \pounds_l\alpha 
         + \quater q^{ca}q^{db}(\pounds_k\gamma_{cd})\pounds_l\gamma_{ab} 
         -\half q^{ab} \pounds_l \pounds_k \gamma_{ab} 
         -\frac{1}{2} \alpha\: q^{ab} \pounds_l\gamma_{ab} 
         -\half q^{ab}\beta_a\beta_b  
\non \\
 &+& 
   \frac{r}{2}\cdot 
    \Bigg[  
          - 2\pounds_l \pounds_l\alpha 
          -\half q^{ab}\pounds_l\gamma_{ab} \cdot
           \left\{ 
                  2\pounds_l\alpha + q^{cd}\beta_c\pounds_l(r\beta_d) 
           \right\} 
\non \\
 && \qquad \, \, 
          - q^{ab}\beta_a\pounds_l\beta_b  
          - \pounds_l\{ q^{ab}\beta_a\pounds_l(r\beta_b)\} 
          - q^{ab}D_a(\pounds_l\beta_b) 
    \Bigg] \,, 
\eena 

\bena
\label{nAR} 
 k^bp^c{}_aR_{bc} &=& 
  - p^b{}_aD_b \alpha 
  + \half\pounds_k\beta_a  
  + \quater \beta_a q^{bc} \pounds_k \gamma_{bc} 
  - p^d{}_{[a}p^e{}_{b]}D_d(q^{bc}\pounds_k \gamma_{ce}) 
\non \\ 
&+& 
   \frac{r}{2}\cdot 
   \Bigg[\; 
         \half (q^{bc}\pounds_k\gamma_{bc})\pounds_l \beta_a 
         + \pounds_k\pounds_l\beta_a  
         + 2\alpha \pounds_l\beta_a  
\non \\ 
 && \qquad 
         + \pounds_l(r\beta_a) \cdot
                       \left\{ 
                              r^{-1}\pounds_l(r^2q^{bc}\beta_b\beta_c) 
                              + 2 \pounds_l\alpha  
                       \right\} 
\non \\
 && \qquad 
         - 2p^b{}_aD_b(\pounds_l\alpha) 
         + \pounds_l(q^{bc}\beta_b\pounds_k\gamma_{ca})  
         - 2r^{-1}\pounds_l \left( 
                                     r^2 q^{cd}\beta_cp^b{}_aD_{[b}\beta_{d]} 
                               \right) 
\non \\
 && \qquad 
         - \half q^{bc}\pounds_l\gamma_{bc}\cdot 
           \Big\{      
                  - (rq^{ef}\beta_e\beta_f+2\alpha)\pounds_l(r\beta_a) 
\non \\ 
&& \qquad \qquad \qquad \qquad \qquad    
                  + 2 p^d{}_aD_d\alpha 
                  - q^{bc}\beta_b\pounds_k \gamma_{ca} 
                  + 2r q^{ef}\beta_ep^d{}_aD_{[d}\beta_{f]} 
           \Big\}  
\non \\
 && \qquad 
         - 2 \pounds_l(\alpha \beta_a) 
         - 2 r(\pounds_l \alpha) \pounds_l \beta_a 
         + p^d{}_aD_b\left\{q^{bc}\beta_c\pounds_l(r\beta_d) \right\} 
\non \\ 
 && \qquad 
         - 2 p^b{}_a q^{cd}D_dD_{[b}\beta_{c]} 
         - q^{bc} (\pounds_l\beta_b) \pounds_k \gamma_{ca} 
\non \\
 && \qquad 
         - q^{bc}\pounds_l(r\beta_b)\cdot 
           \Big\{
                  (rq^{ef}\beta_e\beta_f+2\alpha) \pounds_l\gamma_{ca} 
                  + p^d{}_a D_c \beta_d 
\non \\ 
&& \qquad \qquad \qquad \qquad \qquad  
                  + \beta_c\pounds_l(r\beta_a) 
                  - rq^{ef}\beta_c\beta_f\pounds_l\gamma_{ea}  
           \Big\}     
\non \\
 && \qquad 
          + q^{bc}(\pounds_l\gamma_{ca})\cdot 
             \left\{  
                    2 \beta_b\pounds_l(r\alpha) 
                    + 2 D_b \alpha 
                    - \pounds_k \beta_b   
                    + 2r q^{de}\beta_eD_{[b}\beta_{d]}  
             \right\}  
   \Bigg]\,,  
\eena

\bena
\label{llR} 
 l^a l^b R_{ab} 
        &=& - \half q^{ab} \pounds_l \pounds_l \gamma_{ab} 
            + \quater q^{ca}q^{db}
              (\pounds_l\gamma_{ab})\pounds_l\gamma_{cd} \,,  
\eena

\bena 
\label{lAR} 
 l^bp^c{}_a R_{bc} 
      &=& 
         - \quater \beta_a q^{bc}\pounds_l \gamma_{bc} 
         - \pounds_l \beta_a 
         + \half q^{bc}\beta_c\pounds_l \gamma_{ab} 
         - p^d{}_{[a} p^e{}_{b]}D_d 
           \left(q^{bc}\pounds_l \gamma_{ce} \right) 
\non \\ 
 & + &  \frac{r}{2}\cdot 
        \Bigg[\;  
                - \pounds_l \pounds_l \beta_a 
               + \pounds_l 
                \left(q^{bc}\beta_c\pounds_l\gamma_{ab} \right) 
\non \\
&& \qquad \qquad 
                + \half (q^{cd}\pounds_l \gamma_{cd}) 
                    \left(
                          - \pounds_l\beta_a 
                          + q^{be}\beta_e\pounds_l \gamma_{ab}  
                    \right) 
        \Bigg] \,,  
\eena  

\bena 
\label{ABR}
p^c{}_ap^d{}_b R_{cd} 
        &=& 
         - \pounds_l\pounds_k \gamma_{ab} 
         - \alpha \pounds_l \gamma_{ab} 
         + p^c{}_ap^d{}_b {R}[\gamma]_{cd} 
         - p^c{}_{(a} p^d{}_{b)} D_c\beta_d 
         - \half \beta_a\beta_b 
\non \\  
        &+& 
           q^{cd} \left(\pounds_l\gamma_{d(a}\right)\pounds_k \gamma_{b)c} 
         - \quater 
           \left\{ 
                   (q^{cd}\pounds_k \gamma_{cd})\pounds_l \gamma_{ab} 
                 + (q^{cd}\pounds_l \gamma_{cd})\pounds_k \gamma_{ab} 
           \right\} 
\non \\ 
 &+& \frac{r}{2} \cdot 
     \Bigg[ 
           - 2\alpha \pounds_l \pounds_l \gamma_{ab} 
           - p^e{}_ap^f{}_bD_c(q^{cd}\beta_d\pounds_l \gamma_{ef}) 
\non \\  
  && \qquad \, 
           - \half (q^{cd}\pounds_l\gamma_{cd}) 
             \left\{ 
                     (rq^{ef}\beta_e\beta_f+2\alpha)\pounds_l\gamma_{ab} 
                   + 2 p^e{}_{(a} p^f{}_{b)} D_e\beta_f 
             \right\} 
\non \\ 
  && \qquad \, 
           - 2(\pounds_l \alpha) \pounds_l \gamma_{ab} 
           - r^{-1}\{\pounds_l(r^2q^{ef}\beta_e\beta_f)\} 
                   \pounds_l \gamma_{ab} 
\non \\  
  && \qquad \, 
           - rq^{ef}\beta_e\beta_f \pounds_l \pounds_l \gamma_{ab} 
           - 2\pounds_l \{ p^c{}_{(a} p^d{}_{b)} D_c\beta_d \}  
\non \\ 
  && \qquad \, 
           - 2\beta_{(a}\pounds_l\beta_{b)}
           - r(\pounds_l\beta_a) \pounds_l\beta_b  
           - rq^{ce}q^{df}\beta_c\beta_d 
             (\pounds_l \gamma_{ae})\pounds_l\gamma_{bf} 
\non \\ 
  && \qquad \, 
           + 2q^{cd}\beta_d
             \left\{\pounds_l(r\beta_{(a}) \right\}\pounds_l\gamma_{b)c} 
           + 2p^e{}_{(a}p^f{}_{b)}q^{cd}\left(D_d \beta_e \right) 
             \pounds_l\gamma_{fc} 
\non \\ 
  && \qquad \, 
           + q^{cd} (r q^{ef}\beta_e\beta_f+2\alpha) 
                      (\pounds_l\gamma_{ca})\pounds_l\gamma_{db} \,  
     \Bigg] \,.        
\eena 

\section{Proof of proposition 1}
\label{app:B}

In this proof, we write the Schwarzschild metric in terms of coordinates $(x^A)$ on the $(d-2)$-dimensional symmetry round spheres and coordinates $(x^j)=(t,r)$ parameterizing the 
directions orthogonal to these round spheres. We write $(x^\mu) = (x^j, x^A)$ and $h_{ab} = h_{\mu\nu} (\dd x^\mu)_a (\dd x^\nu)_b$. The Schwarzschild metric 
is written as 
\ben
\bar g = -f(r) \dd t^2 + f(r)^{-1} \dd r^2 + r^2 \dd \Omega^2_{d-2} \equiv g_{ij} \dd x^i \dd x^j + r^2 \Omega_{AB} \dd x^A \dd x^B.
\een
$g_{ij} \dd x^i \dd x^j$ is the metric of the space of $SO(d-1)$-orbits, with derivative $D_i$, and is used to raise and lower indices 
$i,j,k, \dots$.
Following \cite{ishibashi1}, metric perturbations are classified into scalar-, vector- and tensor type as follows. 
\begin{itemize}
\item Scalar perturbations are of the form $h_{ij} = f_{ij} \bS, h_{iA} = rf_i \bS_A, h_{AB} = 2r^2 (H_L \Omega_{AB} \bS + H_T \bS_{AB})$, where $\bS$
denotes a scalar spherical harmonic with eigenvalue $-k^2_S$, where $\bS_A, \bS_{AB}$ are defined as in \cite{ishibashi1}, 
and where $f_i, f_{ij}, H_L, H_T$ do not depend on $x^A$.

\item Vector perturbations are of the form $h_{ij} = 0, h_{iA} = rf_i \bV_A, h_{AB} = -2r^2 k^{-1}_V H_T \hat D_{(A}\bV_{B)}$, where $\bV_A$
denotes a vector spherical harmonic with eigenvalue $-k^2_V$.

\item Tensor perturbations are of the form $h_{ij} = 0, h_{iA} = 0, h_{AB} = 2r^2 H_T \bT_{AB}$, where $\bT_{AB}$. 
denotes a tensor spherical harmonic with eigenvalue $-k_T^2$.
\end{itemize}

The eigenvalues of the spherical harmonics are labelled by $l \in \mathbb{N}_0$, where $l \ge 0$ for scalar-, $l \ge 1$ for vector-, and $l \ge 2$ for tensor harmonics.

For prop. 1, we consider static, asymptotically flat perturbations, i.e. $\partial_t h_{\mu\nu} = 0$, with the usual fall-off conditions for $|h_{\mu\nu}|$ as $r \to \infty$, see def. 2.1 of \cite{CW94} for details. 
The perturbation is additionally required to be 
regular at the horizon, $r=r_0$. \cite{ishibashi, ishibashi1} have constructed gauge invariant master variables for the scalar-, vector- and tensor-type perturbations built from 
$f_i, f_{ij}, H_T, H_L$ and the eigenvalues for each spherical harmonic mode number $l$. In order for static solutions $h_{ab}$ of the described type to exist, one has to find a solution to the appropriate master equation with appropriately regular behavior as $r \to r_0$ and $r \to \infty$. In fact, \cite{ishibashi} have obtained explicit solutions to these equations for the static case for scalar type perturbations in terms of hypergeometric functions of a variable directly related to $r$. They have shown using known asymptotic formulas for hypergeometric functions that no non-zero solution with the prerequisite regularity as $r \to r_0$ and $r \to \infty$ exists, except for a 1-parameter family of perturbations of Schwarzschild corresponding to an infinitesimal change in the Schwarzschild radius $r_0$. By a similar method, one can show that there are no regular, asymptotically flat vector- or tensor type perturbations except for perturbations corresponding to an infinitesimal set of rotation parameters in the $d$-dimensional Myers-Perry family of solutions. We now go through these arguments in detail. Throughout this appendix, it is 
convenient to define 
\ben
n:= d-2, \quad x := \left( \frac{r_0}{r} \right)^{n-1}.
\een
Note that $n$ is not to be confused with the induction order as used in the main text.

\subsection{Tensor perturbations}

For tensor perturbations, the gauge invariant master variable $\Phi$ is defined by $H_T = r^{-n/2}\Phi$. For static perturbations $\Phi = \Phi(x)$ and the master equation is \cite{ishibashi1}
\ben
\begin{split}
&(n-1)^2 x^2 (1-x) \ \Phi''(x) + (n-1)(n+(1-2n)x)x \ \Phi'(x) \\
&-\left( (l-1)(l+n) + \frac{n(n+2)(1-x)}{4} + \frac{n(n+1)}{2x} \right)\ \Phi(x)= 0
\end{split}
\een
The general solution is given in terms of Legendre- and hypergeometric functions by
\ben
\begin{split}
\Phi(x) 
=& \ C_1 x^{-\frac{n}{2(n-1)}} P\left( \frac{-l-n+1}{n-1}, \frac{x-2}{x} \right) + C_2 x^{-\frac{n}{2(n-1)}} Q\left( \frac{-l-n+1}{n-1}, \frac{x-2}{x} \right) \\
=& \ C_3 x^{-\frac{n}{2(n-1)}} F\left( -\frac{l}{n-1}, \frac{l+n-1}{n-1}; 1, \frac{1}{x} \right) + \\
& \ C_4 x^{-\frac{n}{2(n-1)}} \left( \frac{x-2}{x} \right)^{\frac{l}{n-1}} 
\frac{2^{\frac{l}{n-1}} \sqrt{\pi} \Gamma(-\frac{l}{n-1})}{\Gamma(\frac{n-1-2l}{2n-2})} \\
& \times 
F\left( - \frac{l}{2n-2}, \frac{n-1-l}{2n-2}; \frac{n-1-2l}{2n-2}, \frac{x^2}{(x-2)^2} \right)
\end{split}
\een
For asymptotic flatness, we need $h_{AB} = O(r)$ as $r \to \infty$ at a minimum which implies that we should have $\Phi(x) = O(x^{-(n-2)/2(n-1)})$ when $x \to 0$. 
Due to the overall factor, the second term clearly does not satisfy this so we must have $C_4=0$. When $\frac{l}{(n-1)}$ is not a positive integer, 
we may apply a standard linear transformation formula for hypergeometric functions to the first term, resulting in 
\ben
\begin{split}
\Phi(x) =& \ C_3 x^{-\frac{n}{2(n-1)}} \frac{1}{\Gamma(-\frac{l}{n-1})\Gamma(\frac{l+n-1}{n-1})} \sum_{k=0}^\infty \frac{\left(-\frac{l}{n-1} \right)_k \left( \frac{l+n-1}{n-1} \right)_k}{(k!)^2} \\
&\times \left[ 
2\psi(k+1) - \psi\left( -\frac{l}{n-1} +k\right)  - \psi\left( -\frac{l+n-1}{n-1} +k\right) - \log \left( \frac{x-1}{x} \right) 
\right] \left( \frac{1-x}{x} \right)^k
\end{split}
\een
This is singular at the horizon $x=1$ and so is $h_{AB}$. Therefore, we must have $C_3=0$ and thus $\Phi(x)=0$.
 When $\frac{l}{(n-1)}=m$ is a positive integer, the hypergeometric function multiplied by $C_3$ above becomes a polynomial in $1/x$
 of degree at most $m$. Then even for the best possible case $m=1$, $\Phi$ is singular at the horizon and so is $h_{AB}$. Thus, $\Phi(x)=0$ in 
 all cases and we conclude that there cannot exist static, regular, asymptotically flat tensor perturbations. 

\subsection{Vector perturbations}

For vector perturbations, the gauge invariant master variable $\Phi$ is defined in terms of the tensor $F_j$ (eq. 5.10 of \cite{ishibashi1}) by 
$F_i = r^{n-1} \epsilon_{ij} D^j(r^{n/2} \Phi)$, except for the $l=1$ mode. In that case we should instead consider 
the variable $\Phi  =2r \epsilon^{ij} D_{i} (r^{-1} f_{j})$ as the basic gauge invariant potential (eq. 5.16 of \cite{ishibashi1}). For static 
perturbations, $\Phi = \Phi(x)$ in either case.

\noindent{\underline{$l>1$}:} The master equation for a static vector perturbation is 
\ben
\begin{split}
&(n-1)^2 x^2 (1-x) \ \Phi''(x) + (n-1)(n+(1-2n)x)x \ \Phi'(x) - \\
&-\left( (l-1) (l+n) + \frac{n(n+2)}{4} - \frac{3n^2 x}{4} \right)\  \Phi(x) = 0
\end{split}
\een
The general solution is
\ben
\begin{split}
\Phi(x) =& \ C_1 x^{-\frac{2l+n}{2(n-1)}} F\left(\frac{-n-l}{n-1}, \frac{n-l}{n-1}; -\frac{2l}{n-1}, x\right) + \\
& \ C_2 x^{\frac{2l+n-2}{2(n-1)}}F\left(  \frac{l-1}{n-1}, \frac{l+2n-1}{n-1}; \frac{2l+2n-2}{n-1}, x\right)
\end{split}
\een
At large distances, $x \to 0$, the terms behave as $\sim C_1 x^{-\frac{2l+n}{2(n-1)}}$ respectively as $\sim C_2x^{\frac{2l+n-2}{2(n-1)}}$. One sees 
from this that for the perturbation to be asymptotically flat we must require $C_1=0$. By applying a linear transformation formula for 
hypergeometric functions to the second linearly independent solution multiplied by $C_2$, we see that 
\ben
\begin{split}
\Phi(x) =& \ C_2 x^{\frac{2l+n-2}{2(n-1)}} 
\frac{\Gamma(\frac{2l+2n-2}{n-1})}{\Gamma(\frac{l-1}{n-1}) \Gamma(\frac{l+2n-1}{n-1})} \sum_{k=0}^\infty   \frac{\left(\frac{l-1}{n-1} \right)_k \left( \frac{l+2n-1}{n-1} \right)_k}{(k!)^2} \\
&\times \left[ 
2\psi(k+1) - \psi\left( \frac{l-1}{n-1} +k\right)  - \psi\left( \frac{l+2n-1}{n-1} +k\right) - \log \left( 1-x \right) 
\right] \left(1-x \right)^k
\end{split}
\een
This has a logarithmic divergence for $x \to 1$, i.e. at the horizon. 

\noindent{\underline{$l=1$}:} For static perturbations, the linearized Einstein equation yields $\Phi(r) = L/r^{n+1}$ for some real constant $L$, from which we learn that the 
only nonzero component of $h_{\mu\nu}$ is 
$h_{At} = r f_t \bV_A = \frac{L}{(n+1)r^{n-1}} \bV_A$, where $\bV^A$ is a Killing vector field of the $n$-sphere. Such a perturbation corresponds to turning on an infinitesimal 
rotation parameter in the Myers-Perry solution. 

To summarize, the only static, regular, asymptotically flat vector perturbations correspond to adding perturbatively a rotation to the Schwarzschild black hole towards
a Myers-Perry black hole. 

\subsection{Scalar perturbations}

For scalar perturbations, we take as our gauge invariant 
master variable, $Y$, the combination\footnote{Its relation to the master variable $\tilde Y$ obeying the Schr\" odinger-type equation 3.6 of \cite{ishibashi} is $\tilde Y = \sqrt{\frac{f}{r}} \frac{r}{f'} Y$.} 
$Y= r^{n-2}(f^{-1} F^{ij} (D_ir)D_jr -2F)$, where $F_{ij},F$ are defined in eqs. 2.7a,b of \cite{ishibashi1}. For static scalar perturbations, $Y=Y(x)$, which is a solution to the master equation
\ben
\begin{split}
&(n-1)^2 x^2 (1-x) \ Y''(x) - 2(n-1)(1+(n-2)x)x \ Y'(x) + \\
&+\bigg( (n-2)x - (l-1)(n+l) \bigg)\  Y(x) = 0
\end{split}
\een
The general solution is
\ben
\begin{split}
Y(x) =& \ C_1 x^{-\frac{l-1}{n-1}} F\left(\frac{n-l-1}{n-1}, -\frac{l}{n-1}; -\frac{2l}{n-1}, x\right) + \\
& \ C_2 x^{\frac{l+n}{n-1}}F\left(  \frac{n+l-1}{n-1}, \frac{2(n-1)+l}{n-1}; \frac{2(l+n-1)}{n-1}, x\right)
\end{split}
\een
To see which of these solutions correspond to a regular, asymptotically flat pertrubation, we apply a gauge transformation to the perturbation to bring it into Regge-Wheeler
gauge in which $f_i = H_T = 0$. The transformed perturbation is still static, regular, and asymptotically flat and takes the form 
\ben
h_{\mu\nu} \dd x^\mu \dd x^\nu = -F_t^t f \bS \  \dd t^2 + f^{-1} F_r^r \bS \ \dd r^2 + 2Fr^2 \bS \ \Omega_{AB} \dd x^A \dd x^B
\een
where in the coordinate $x$ the metric function $f(x) = 1-x$ and the horizon is at $x=1$. The components of the gauge invariant variables $F_i^j, F$ are given 
by the master variable $Y$ and $X=Y+2fY'/f'$ through the relations
\ben
F^t_t = \frac{(n-1)X-Y}{nr^{n-2}}, \quad F_r^r = \frac{-X+(n-1)Y}{nr^{n-2}}, \quad F_t^r = 0, \quad F= -\frac{X+Y}{2nr^{n-2}},
\een
noting that $Z=0$, see eqs. 4.2 of \cite{ishibashi}. 

\noindent{\underline{$l>0$}:}
Using this together with the asymptotic forms of the hypergeometric functions, we see that the first 
solution multiplied by $C_1$ would give rise to $F^t_t, F_r^r, F$ diverging as $r^l$ for $r \to \infty$. Therefore, asymptotic flatness requires $C_1=0$.
By applying an appropriate linear transformation formula to the hypergeometric function multiplied by $C_2$, we obtain
\ben
\begin{split}
Y(x) =& \ C_2 x^{\frac{n+l}{n-1}} \Bigg[
\frac{\Gamma(\gamma_2)}{\Gamma(\alpha_2)\Gamma(\beta_2)} (1-x)^{-1} + \\
& \ + \frac{\Gamma(\gamma_2)}{\Gamma(\alpha_2-1)\Gamma(\beta_2-1)}
\sum_{k=0}^\infty   \frac{\left(\alpha_2\right)_k \left( \beta_2 \right)_k}{k!(k+1)!} 
 \{
\log(1-x) - \psi(k+1) - \psi(k+2) + \\
& \ \hspace{3cm}  \psi(\alpha_2+k) + \psi(\beta_2+k)
\} (1-x)^k
\Bigg]
\end{split}
\een
where $\alpha_2 = \frac{n+l-1}{n-1}, \beta_2 = \frac{2(n-1)+l}{n-1}, \gamma_2 = \alpha_2 + \beta_2 -1$. Note that $\alpha_2>1,\beta_2>1$. Near the horizon
we therefore find that for $x \to 1$ we have the asymptotic behavior
\ben
X \sim -Y \sim - C_2 
\frac{\Gamma(\gamma_2)}{\Gamma(\alpha_2)\Gamma(\beta_2)} (1-x)^{-1}
\een
and therefore $F^t_t$ and $F^r_r$ both diverge as $f(r)^{-1}$ at the horizon $r \to r_0$. Thus, there are no static, regular, asymptotically flat scalar type perturbations
when $l>0$. 

\noindent{\underline{$l=0$}:} In this case, we have a spherically symmetric, static perturbation. It can be shown by analyzing the perturbation equations that these 
correspond up to gauge precisely to a perturbative change in the Schwarzschild radius, $r_0$, hence to a perturbation towards another Schwarzschild black hole. 

\medskip

Combining the results for scalar-, vector- and tensor perturbations and taking into account the gauge transformations implicitly considered in the above arguments, we 
have shown that $h_{\mu\nu} = z_{\mu\nu} + \pounds_\xi \bar g_{\mu\nu}$, where $z_{\mu\nu}$ corresponds to an infinitesimal variation of the parameters in the Myers-Perry 
solution \cite{mp}. By assumption the coordinate components $h_{\mu\nu}$ in an asymptotically Cartesian coordinate system built from $(t,r,x^A)$ fulfill the 
fall-off conditions detailed in \cite{CW94}, def. 2.1, for $r \to \infty$. Since furthermore $\partial_t h_{\mu\nu} = 0 = \pounds_t z_{\mu\nu}$, it follows that $\pounds_\xi \bar g_{\mu\nu}$
is asymptotically flat at $\cI^\pm$, as we can see e.g. by analyzing its behavior in the coordinates $(t \pm r, r, x^A)$. Thus, $\xi^a$ is an asymptotic symmetry 
at $\cI^\pm$.

\qed

\end{document}